\documentclass[journal]{IEEEtran}

\usepackage{amsmath,amsfonts,amssymb,bm}
\usepackage{algorithmic}
\usepackage{algorithm}
\usepackage{graphicx}
\usepackage{subfig}
\usepackage{textcomp}
\usepackage{xcolor}
\usepackage{svg}
\usepackage{extarrows}

\usepackage{epstopdf}
\usepackage{enumerate}
\usepackage[T1]{fontenc}

\usepackage{float}
\usepackage{svg}

\usepackage{multirow}

\usepackage{textcomp}

\usepackage{siunitx}
\usepackage{cases}
\usepackage{extarrows}

\begin{document}

\title{SFL-LEO: Asynchronous Split-Federated Learning Design for LEO Satellite-Ground Network Framework}

\author{Jiasheng Wu, Jingjing Zhang~\IEEEmembership{Member,~IEEE}, Zheng Lin, Zhe Chen~\IEEEmembership{Member,~IEEE}, Xiong Wang~\IEEEmembership{Member,~IEEE}, Wenjun Zhu, Yue Gao~\IEEEmembership{Fellow,~IEEE}

\thanks{J. Wu, Z. Lin, Z. Chen, X. Wang, W. Zhu, and Y. Gao are with the Institute of Space Internet, Fudan University, Shanghai 200438, China, and the School of Computer Science, Fudan University, Shanghai 200438, China.}
\thanks{J. Zhang is with the Institute of Space Internet, Fudan University, Shanghai 200438, China, and the School of Information Science and Technology, Fudan University, Shanghai, China.}
}


\maketitle

\begin{abstract}
Recently, the rapid development of LEO satellite networks spurs another widespread concern---data processing at satellites. However, achieving efficient computation at LEO satellites in highly dynamic satellite networks is challenging and remains an open problem when considering the constrained computation capability of LEO satellites.  
For the first time, we propose a novel distributed learning framework named SFL-LEO by combining Federated Learning (FL) with Split Learning (SL) to accommodate the high dynamics of LEO satellite networks and the constrained computation capability of LEO satellites by leveraging the periodical orbit traveling feature. The proposed scheme allows training locally by introducing an asynchronous training strategy, i.e., achieving local update when LEO satellites disconnect with the ground station, to provide much more training space and thus increase the training performance. Meanwhile, it aggregates client-side sub-models at the ground station and then distributes them to LEO satellites by borrowing the idea from the federated learning scheme. Experiment results driven by satellite-ground bandwidth measured in Starlink demonstrate that SFL-LEO provides a similar accuracy performance with the conventional SL scheme because it can perform local training even within the disconnection duration.
\end{abstract}


\IEEEpeerreviewmaketitle

\section{Introduction}\label{sec:introduction} 
Nowadays, we are witnessing the explosive growth of a new kind of Internet service offered by the Low Earth Orbit (LEO) satellite networks such as Starlink and Oneweb \cite{oneweb,starlink,lin2025esl,yuan2025constructing}, which is also an important part in future 6G networks. As specified in \cite{darwish2022leo,6G-road,lin2025leo,yuan2024satsense}, terrestrial and non-terrestrial networks based on LEO satellite networks constitute the future 6G networks in a complementary way~\cite{liu2024efficient,peng2024sums,yuan2023graph}. Specifically, the non-terrestrial networks can provide worldwide coverage which can not be offered by the conventional terrestrial networks such as LTE and 5G networks\cite{zhao2024leo,zhang2024satfed,peng2025sigchord}. In the context of future 6G networks featured with integrated space-air-ground network framework, abundant computation resource concealed in large LEO satellite constellations provides an appealing option for mobile edge computation at LEO satellites, which can bring several side benefits like alleviating the pressure of redundant data delivery from satellites to the ground.  

However, it is non-trivial to achieve efficient computation in LEO satellite networks due to the high dynamics of LEO satellites and the constrained computation capability of single LEO satellite~\cite{lin2024fedsn,zhang2025s}. For example, compared to traditional geostationary satellites, LEO satellites travel at a high speed of around 7.9 km/s in relative to the ground, which results in a limited connectivity period (e.g., around 3-9 minutes) with the ground station. It should be noted that the disconnection state accounts for the majority of time when orbiting the Earth. Consequently, there exists limited time for data and task offloading from LEO satellites to the ground station. To overcome this challenge, recently there exists a body of research work on traffic scheduling from LEO satellites to the ground, in order to reduce the delay of data offloading \cite{carvalho2019optimizing,castaing2014scheduling}. However, these work remains inefficient when downloading massive onboard data to the ground due to orbital dynamics and bandwidth constraints. Meanwhile, there exists another line of research work on data compression at LEO satellites and thus reducing the amount of data delivered from LEO satellites to the ground station \cite{orbital2020denby}, which poses a heavy pressure on resource-constrained satellites. To accommodate for the limited resource in LEO satellites, offloading partial data processing tasks to the ground station with much higher computation capability is a more optimal option than \cite{orbital2020denby}.

  
Therefore, we propose a novel Split-Federated Learning (SFL) design named SFL-LEO for LEO satellite-ground network topology by making use of the computation resource at ground stations. The advantages brought by this design are two-fold. First, it can significantly alleviate the computation pressure on LEO satellites. Secondly, the training speed is accelerated. Specifically, we divide data training tasks into two parts, where one part is assigned to LEO satellites according to their computation capability while the remaining tasks are delivered to the ground station. At the heart of this design is an asynchronous split-federated learning scheme leveraging the periodical orbit travelling. For the first time, we demonstrate how to combine the SL scheme~\cite{thapa2022splitfed,lin2024adaptsfl,lyu2023optimal,lin2024efficient} with FL scheme in a complementary way and then applied for data processing in satellite networks. Due to the constrained computation capability, we cannot directly leverage the FL scheme, which is beyond the ability of LEO satellites. Alternatively, we design an adaptive SL strategy, which enables task assignments that match with individual satellite computation capability, to achieve heterogeneous computation in LEO satellite networks. However, due to the inability to cover the characteristics of global data, there exists a limited generalization ability with the SL model trained by each satellite themselves. Hence, we employ a modified FL scheme to aggregate the models trained by different satellites at the ground station and then distribute to these LEO satellites, in order to improve the generalization ability and training accuracy.


However, the implementation of the proposed design in LEO satellite networks entails two challenges. First, during one orbit period of around 100 minutes \cite{InvestingStarlink}, the short-duration connectivity (e.g., usually less than 9 minutes) and long-duration disconnection between LEO satellites and the ground station restricts the trained model size on satellites due to the short connectivity duration and constrained computation capability, thus rendering the conventional SL scheme unsuitable for the LEO satellite-ground network framework since it requires maintaining the link connectivity between the ground station and LEO satellites during the training process. Secondly, the fixed splitting strategy also disables the conventional SL scheme in this case since there exists heterogeneous computation capability within the LEO satellites, which calls for an adaptive splitting scheme. Meanwhile, only limited generalization can be acquired with the SL model trained by each satellite themselves, which thus decreases the overall training performance.


To deal with the first challenge, we devise a novel local update scheme which can perform training locally at LEO satellites even when LEO satellites disconnect with the ground station. In particular, we introduce an auxiliary network for each satellite. Leveraging the auxiliary network, offline training can be performed on LEO satellites and then update splitting network and auxiliary network during the disconnection period. Upon building the connection with the ground station, LEO satellites transmit the smashed data and training parameters of the splitting network to the ground station for aggregation. In this way, the amount of data delivered from satellites to the ground station can be reduced, as well as the delay. To resolve the second challenge, we propose an adaptive split learning model which assigns partial training tasks to LEO satellites based on their computation capability and then assign the remaining training tasks to the ground station. However, it incurs another problem--different sizes of splitting network assigned for LEO satellites, rendering model aggregation and generalization at the ground station infeasible. To deal with this problem, we have proposed a novel splitting network layer alignment strategy, the core idea of which is to make sure that the size of last layer in the splitting network of all LEO satellites remains the same by intelligently removing certain intermediate layers. Consequently, splitting networks with different sizes from different LEO satellites can be aggregated at the ground station and then SFL-LEO can accommodate for resource-constrained and computation-heterogeneity LEO satellites. 


Finally, we have built a prototype of 5G Non-Terrestrial-Network using Open5GS for simulating the LEO satellite-ground network topology, which is driven by real Starlink traces and measured bandwidth between satellites and ground station. Based on this prototype, we have conducted extensive experiments using remote sensing image dataset. Experiment results demonstrate that the data amount from satellites to the ground based on the proposed scheme can be decreased by 6$\times$ compared to the centre scheme. Here, the centre scheme means that all data are transmitted to the ground station for training. Meanwhile, the data amount can be reduced by 15$\times$ and 60$\times$ in comparison with the SL and FL schemes, respectively. Actually, the training accuracy based SFL-LEO is around 15\% and 16\% higher than the conventional SL and FL schemes, respectively. However, SFL-LEO demonstrates a worse training performance (around 7\% lower) compared to the centre scheme since all data can be trained simultaneously in the centre scheme. Actually, SFL-LEO is a design that makes a tradeoff between the amount of transmitted data and training accuracy. The contribution of this paper can be summarized as follows.

\begin{itemize}
\item We present a novel hybrid split-federated learning framework for LEO satellite networks by integrating the SL scheme with the FL scheme in a complementary way. To the best of our knowledge, this work represents the poineering research efforts to bring distributed training to LEO satellite systems, opening up new possibilities for data processing in this context. 

\item To overcome the high dynamics and computation-heterogeneity of LEO satellites, we propose a local update scheme operating upon the designed auxiliary network to tackle the intermittent connectivity issue, and meanwhile design a training network layer alignment strategy to accommodate for different splitting network sizes assigned for LEO satellites in the LEO satellite-ground topology. 


\item To verify the effectiveness of our approach, we develop a prototype and conduct extensive experiments driven by real Starlink traces and satellite-ground bandwidth. Experiment results based on remote sensing image data provide empirical evidence of advantages of our proposed framework compared to existing approaches.
\end{itemize}

The rest of this paper is organized as follows. Section II presents 
the system model including the satellite model and the split learning architecture. Section III introduces the proposed communication-efficient asynchronous split learning algorithm. In Section IV,
experiments are conducted to evaluate the performance of our
method. Finally, Section V concludes this paper.

\section{Background and Motivation}

In this section, we commence by introducing the satellite-ground communication network. Building upon this foundation, we unveil an innovative split-federated framework that facilitates collaborative  model training between LEO satellites and the ground station. Subsequently, we demonstrate why the SFL framework cannot be directly applied in the satellite-ground network topology, motivating the proposed design.


\subsection{Background for Satellite Constellation} \label{back1}
\begin{figure}[t!]
\centering
   \subfloat{\includegraphics[width=0.4\textwidth]{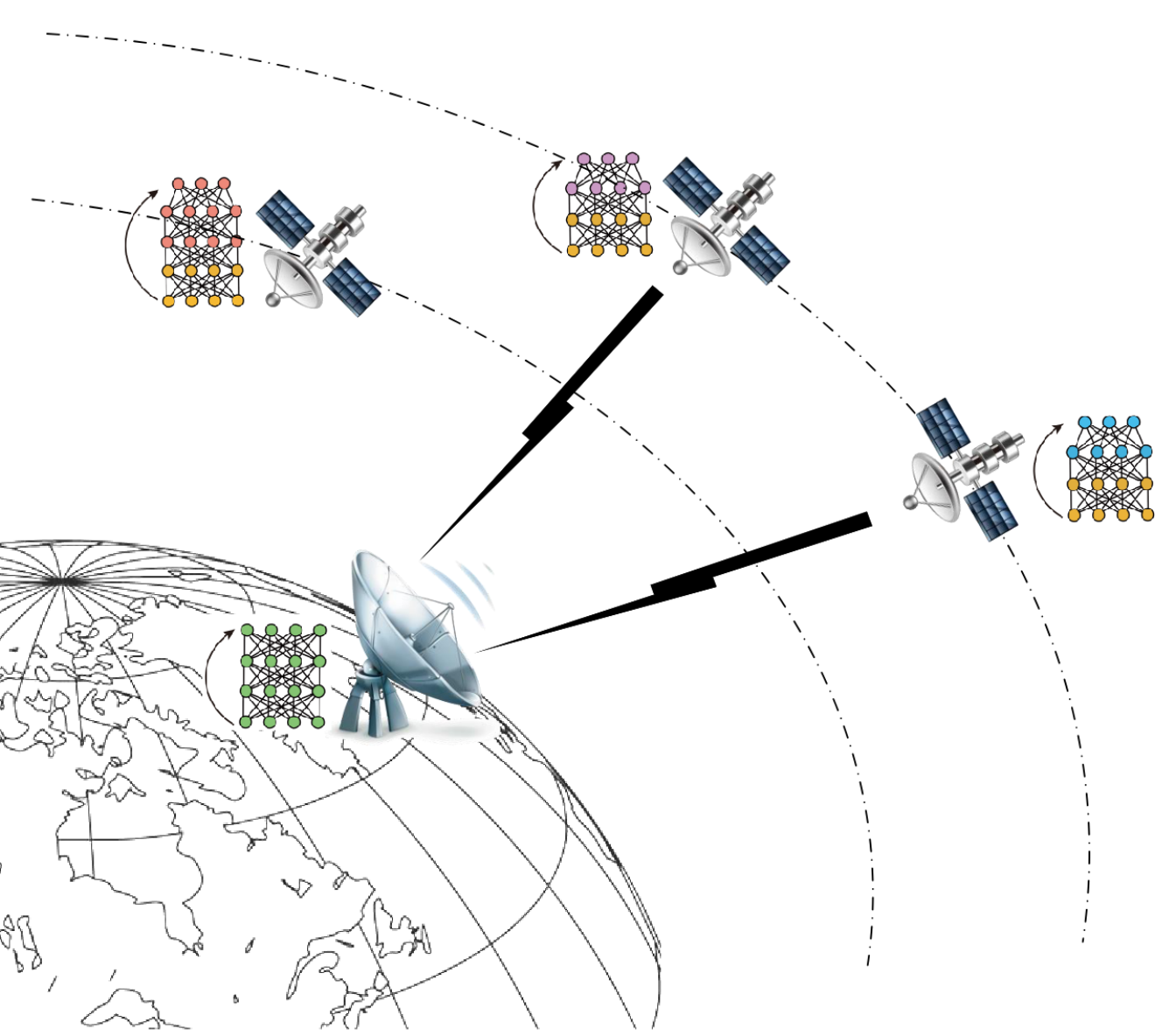}\label{1}}
    \caption{(a) A uniform constellation of $N=3$ satellites in $m=2$ orbits at two different altitudes.}\label{fig:SA-GS}
\end{figure}

As illustrated in Fig. \ref{fig:SA-GS}, we delve into the investigation of a satellite-ground communication model. It is composed of a constellation with $N$ satellites (e.g., the satellite set $\mathcal{N}=\{1, 2, \cdots, N\}$) operating within $m$ orbital planes, and connect with a single ground station (GS). The flexible constellation configuration such as inclination and orbit provides a range of available options, for example, from the renowned Walker constellation \cite{walker1984satellite} to the groundbreaking SpaceX constellation \cite{del2019technical}.

In an earth-centered inertial coordinate system, each satellite $n \in \mathcal{N}$ follows a distinct 3D trajectory $\mathbf{r}_n(t)$, while the GS itself follows a 3D trajectory $\mathbf{r}_g(t)$, with $t$ denoting real-world time. Each satellite $n$ establishes a connection with the GS only when it is within the GS's line of sight range. This condition can be satisfied when the angle between the line of sight from the GS to the satellite and the local vertical direction at the GS is greater than or equal to $\alpha_{\text{min}}$. Mathematically, this condition can be expressed as:
\begin{equation}
\alpha_{n,g} = \angle\left({\bf{r}}_{g}(t),{\bf{r}}_{n}(t) - {\bf{r}}_{g}(t)\right) \leq \frac{\pi}{2} - \alpha_{min},
\end{equation}
where $\alpha_{\text{min}}$ represents the minimum elevation angle for building the connection between satellites and the ground station in LEO satellite networks. The above requirement is set to ensure that the satellite is positioned above the prescribed threshold relative to the GS, allowing for data delivery between the satellite and ground station.

\begin{figure}[t!]
\centering
\includegraphics[width=3.2in]{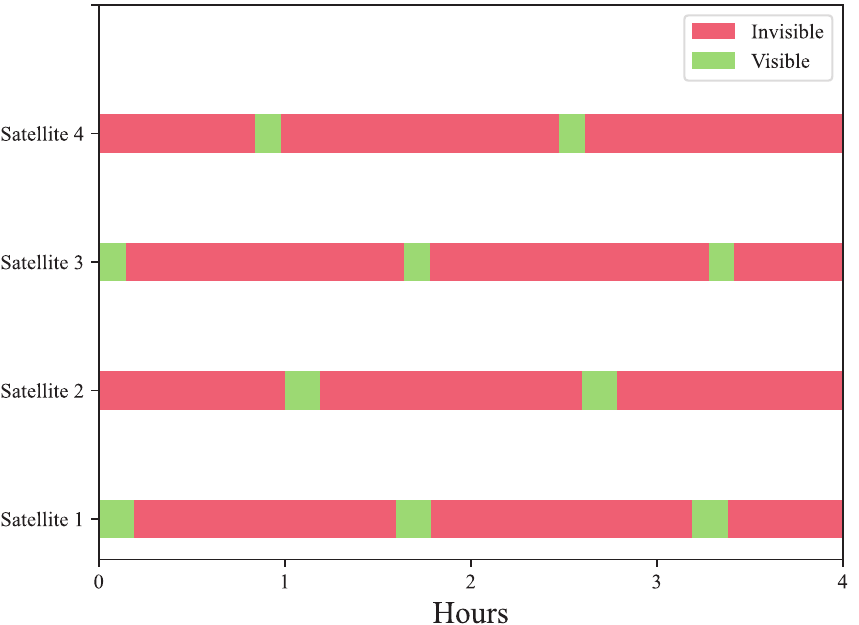}
\caption{Visible patterns of satellites. Satellites 1 and 2 are at an altitude of $550$ Km, Satellites 3 and 4 are at an altitude of $340$ Km, and the GS is at the North Pole.} 
\label{fig:SA_period}
\end{figure}

LEO satellites usually experience much shorter visibility period with the GS compared to the whole orbit period, as as demonstrated in Fig. \ref{fig:SA_period}. As a result, we can observe short-duration link establishment, yet long-duration disconnection between LEO satellites and the ground station, thereby introducing the communication staleness that can hinder the training convergence. In other words, the information returned by each satellite is computed at a stale value of the global parameter. Hence, it is imperative for us to reevaluate the disparities in the local models of the satellites for FL aggregation. Moreover, satellites at higher altitudes revisit the GS less frequently compared to those orbiting at lower altitudes, resulting in longer disconnection period and slower convergence rate in the synchronous distributed learning model. Therefore, SFL-LEO with asynchronous communication mode emerges as a more suitable approach for satellite-ground communication scenarios, offering the potential to alleviate the adverse effects of intermittent communication.


\subsection{Background on SFL Framework in LEO Satellite Networks}\label{sec:AFL}

The rise of satellites offering diverse functionalities has spurred significant growth in satellite payloads. This expansion, in response, necessitates the incorporation of onboard computing capabilities to address a multitude of tasks, spanning from image processing and pattern recognition to space modulation recognition. However, as the data volume collected by each satellite undergoes exponential growth, we have implemented the SFL framework within LEO satellite networks to ensure the communication-efficient execution of these computational tasks.

We now proceed to present the SFL framework for LEO satellite networks, where the $N$ satellites function as clients, each with its own dataset, while the single GS operates as the server.
The objective of the training process is to minimize the global loss function given as 
\begin{equation}  \label{eq:obj}
\min_{{\bf{w}}\in \mathbb{R}^{d}}\ F({\bf{w}}) 
:= \min_{{\bf{w}}\in \mathbb{R}^{d}} \frac{1}{N} \sum_{n=1}^{N} p_n F_n\left({\bf{w}}\right),
\end{equation}
where $\bf w$ represents the model parameter to be optimized; $F_n({\bf{w}})$ quantifies the average loss of ${\bf{w}}$ on the local dataset $\mathcal{D}_n$ of the $n$-th satellite, which consists of $d_n$ samples; and $p_n = d_n/(\sum_{n=1}^{N} d_n)$ weights the importance of $\mathcal{D}_{n}$.


To enable SFL, the global model parameter $\mathbf{W}$ is divided into two components, denoted as $\mathbf{w}=[\mathbf{w}_s,\mathbf{w}_g]$. Here, $\mathbf{w}_{s}$ represents the client-side model computed at the satellites, while $\mathbf{w}_g$ represents the server-side model computed at the GS. Consequently, this framework incorporates customized loss functions for both the clients and the GS.

\textbf{Satellites}: To facilitate local updates, an auxiliary network $\mathbf{a}_s$ is incorporated into the satellites to compute the local loss []. In this setup, the auxiliary network $\mathbf{a}_s$ consists of multiple layers connected to the satellite model ${\bf{w}}_s$, where the output of ${\bf{w}}_s$ serves as the input for $\mathbf{a}_s$. 

For the satellites, the goal is to find the optimal parameters ${\bf{w}}_s^*$ and $\mathbf{a}_s^*$ that minimize the loss function $F_s\left({\bf{w}}_s\right)$, given as 
 \begin{equation}  \label{eq:local}
\min_{{\bf{w}}_s, \mathbf{a}_s}\ F_s({\bf{w}}_s) := \min_{{\bf{w}}_{s}, \mathbf{a}_s} \frac{1}{N} \sum_{n=1}^{N} F_{s,n}({\bf{w}}_{s}, \mathbf{a}_s) ,
\end{equation}
where $F_{s,n}({\bf{w}}_{s}, \mathbf{a}_s)$ is the local loss function of the current model at satellite $n$, defined as $F_{s,n}({\bf{w}}_{s}, \mathbf{a}_s)= \frac{1}{d_n}\sum_{x \in \mathcal{D}_{n}} l_n\left(x,{\bf{w}}_{s}, \mathbf{a}_s \right)$. Here, $l_n\left(x,{\bf{w}}_{s}, \mathbf{a}_s \right)$ is the loss function evaluated with input $x$, model ${\bf{w}}_{s}$ and the auxiliary model $\mathbf{a}_s$. 


By incorporating the auxiliary network, the satellites can perform local updates without the need to wait for the transmission of the gradients of the smashed data. 

\textbf{GS}: Based on the optimal model ${\bf{w}}_s^*$, the goal of GS is to find the optimal model ${\bf{w}}_g^*$ that minimizes the loss function $F_g\left({\bf{w}}_g\right)$, given as 
 \begin{equation}  \label{eq:global}
\min_{{\bf{w}}_g}\ F_g({\bf{w}}_g) 
= \min_{{\bf{w}}_g} \frac{1}{N} \sum_{n=1}^{N} F_{g,n}({\bf{w}}_g, {\bf{w}}_s^*) ,
\end{equation}
where $F_{g,n}({\bf{w}}_g, {\bf{w}}_s^*)$ is the local loss function of the current GS-side 
model corresponding to satellite $n$, defined as $F_{g,n}({\bf{w}}_g, {\bf{w}}_s^*)= \frac{1}
{d_n}\sum_{{\bf{x}} \in \mathcal{D}_{n}} {l\left( {\bf{w}}_g, g_{{\bf{w}}_s^*} (x) \right)}$. Here, 
note that the smashed data of the optimal model ${\bf{w}}_s^*$ given the input $x$ is denoted as $g_{{\bf{w}}_s^*}(x)$.

To expedite the training process, an asynchronous stochastic gradient descent (SGD) approach with local updates is employed. Specifically, after the GS aggregates the smashed data from $K \leq N$ satellites sequentially based on the constellation, it generates a global model and transmits it to the following $K$ satellites for the next round of training. 

\subsection{Heterogeneity Enforces Model Split Personalization} \label{back2}


LEO satellites play a pivotal role in various domains, including remote sensing, weather monitoring, and cutting-edge scientific research. Each distinct mission objective necessitates a bespoke blend of computational resources and capabilities, tailored to its unique demands. Furthermore, the onboard processors embedded within LEO satellites exhibit a range of attributes, encompassing processing power, architectural intricacies, and operational speed. This diversity in processor characteristics arises from the intricate interplay of mission objectives, technological designs, and the availability of resources.

Consequently, the landscape of LEO satellites unfolds with notable disparities in computational prowess. These variations, rooted in the convergence of multifaceted mission goals, technological trajectories, and resource allocations, paint a dynamic canvas where each satellite emerges with a distinct computational profile. This interweaving of diverse factors cultivates a tapestry of heterogeneity that permeates the realm of LEO satellites. It necessitates a strategic approach to model segmentation, one that embraces precision and personalization to accommodate this nuanced diversity effectively.

\section{Design Overview}

In this work, we propose SFL-LEO---a novel distributed learning framework designed to cater to the dynamic nature of LEO satellite networks and the limited computational capabilities of LEO satellites. SFL-LEO enables satellites to engage in learning during periods of invisibility while also employing a personalized model split strategy.

\begin{figure}[t!]
\centering
\includegraphics[width=3.3in]{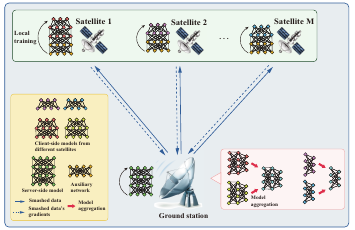}
\caption{Design Overview.} 
\label{fig:Design Overview}
\end{figure}

Fig. \ref{fig:Design Overview} reveals the architecture of the proposed SFL-LEO framework. The framework can be primarily divided into two main components: satellite and GS. On the satellite side, each individual satellite trains a personalized split network configuration. The satellites then transmit their parameters when they are connected. On the GS side, which possesses robust computational capabilities, it undertakes the primary role in network computations. Each connection updates itself based on the data received from the satellites.

In our framework, a mechanism to support local updates for satellites has been designed. By introducing an auxiliary network, satellites can perform local updates during disconnection from the GS and engage in global aggregation upon reconnection. This strategy enables efficient utilization of satellite disconnection periods for training purposes, which resolves the challenge posed by long periods of satellite invisibility. This mechanism also effectively reduces the volume of traffic transmitted through the satellite-to-ground link during network training, thereby alleviating the traffic pressure on the satellite-to-ground link.

Another distinguishing design of the proposed framework lies in the strategy of network splitting. Traditional splitting strategies, when multiple network sizes coexist, often lead to a significant drop in network performance. In contrast, we ensure the consistency of the cutting layers while omitting hidden layers in the split network. This approach effectively preserves the overall performance when different-sized splitting networks coexist. Consequently, we are capable of deploying networks of varying sizes on different satellites, thus making efficient use of their computational capabilities.

Acknowledging the operational characteristics of satellites, the GS incorporates the concept of staleness during the aggregation process. Based on the number of updates the satellite has undergone since its last connection with the ground station, the GS evaluates the model's freshness and applies corresponding weights to the model from that satellite. Furthermore, we are exploring the stability of the connections. In cases where a connection cannot be established due to factors like adverse weather conditions or physical obstructions, it results in an increase in staleness. To address this issue, we are considering the implementation of penalties based on staleness. These designs ensure effectiveness for satellites operating within the same cycle as well as those operating on different cycles.


\section{SFL-LEO's Design} 

In this section, we introduce a novel communication-efficient asynchronous SFL approach for LEO satellite networks. The approach hinges on a tripartite concept. Firstly, to perform local computations, we capitalize on the windows when communication is not feasible, known as the ``invisible'' time of the satellites. This maneuver maximizes resource utilization during these temporal gaps. Secondly, we partition the model based on the computing capacity of the satellites. This division enables us to optimize the utilization of computing resources in a distributed manner, allocating tasks according to each satellite's capabilities. Lastly, we adopt an asynchronous mechanism that accounts for the potential staleness of local gradients by integrating a penalty into the aggregation process. Through the orchestration of this dual-pronged strategy, our objective is to markedly elevate the overall efficiency and efficacy of the SFL framework within the realm of LEO satellite networks.

To elucidate the proposed approach more comprehensively, we commence by illustrating the scenario in which the satellites possess akin computing capabilities. Subsequently, we expand our discussion to encompass the broader context that accounts for varying degrees of computational prowess among the satellites.

\begin{figure}[t!]
\centering
\includegraphics[width=2.5in]{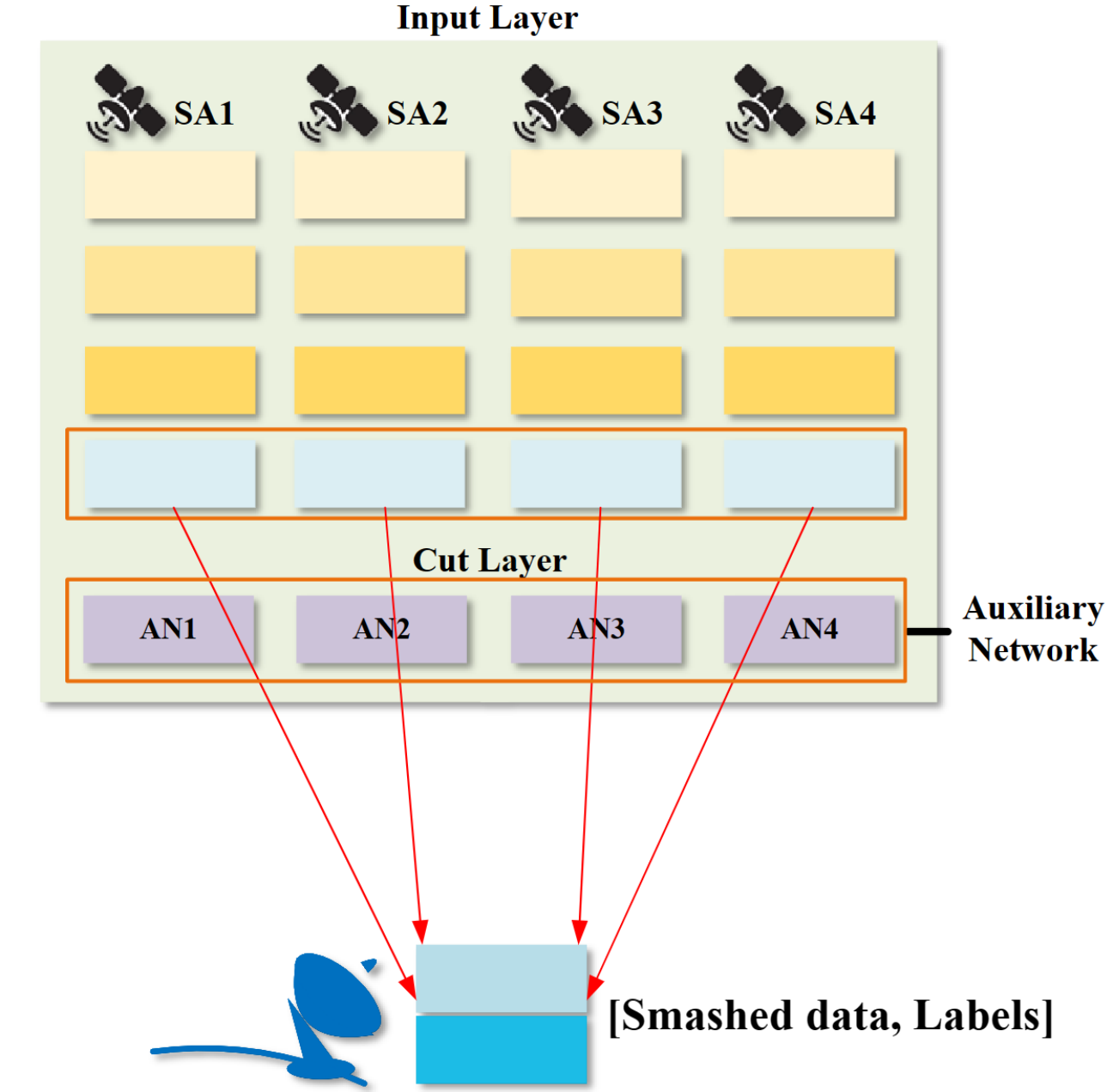}
\caption{The satellites have comparable computing capabilities and have the same split strategy.}
\label{fig:plot_abstract}
\end{figure}

\subsection{Combining FL and SL  for  LEO Satellite Networks}
In this subsection, we present the benefits of fusing SL and FL in the satellite scenarios. Our approach seamlessly integrates aspects of both FL and SL to create an effective and adaptive learning framework tailored to LEO satellite networks. We achieve this through an asynchronous split-federated learning scheme that maximizes training efficiency and performance in the presence of intermittent satellite-ground connections.

Federated Learning, by design, fosters decentralized training, allowing models to be trained on local data without centralizing it. This suits the asynchronous nature of LEO satellite communication where satellites might often lose connection with the ground station. On the other hand, Split Learning is renowned for dividing the deep learning model into segments, where initial layers can be trained locally, and the latter layers are trained at a centralized server, reducing the computational burden on devices with limited capability. This is a boon for LEO satellites with their computational constraints. SFL-LEO's asynchronous split-federated learning mechanism aligns perfectly with the unique features of LEO satellite networks. The intermittent disconnection periods are effectively utilized for local training, mitigating the impact of communication constraints. The adaptive split learning strategy optimizes resource allocation, while the aggregation mechanism guarantees consistent model quality.

\subsection{Identical Model Split}
We commence by presenting a straightforward scenario where all satellites possess comparable computing capabilities. In this instance, we employ an identical model split, wherein each satellite computes the same local model.

\begin{figure}[t!]
\centering
\includegraphics[width=3.4in]{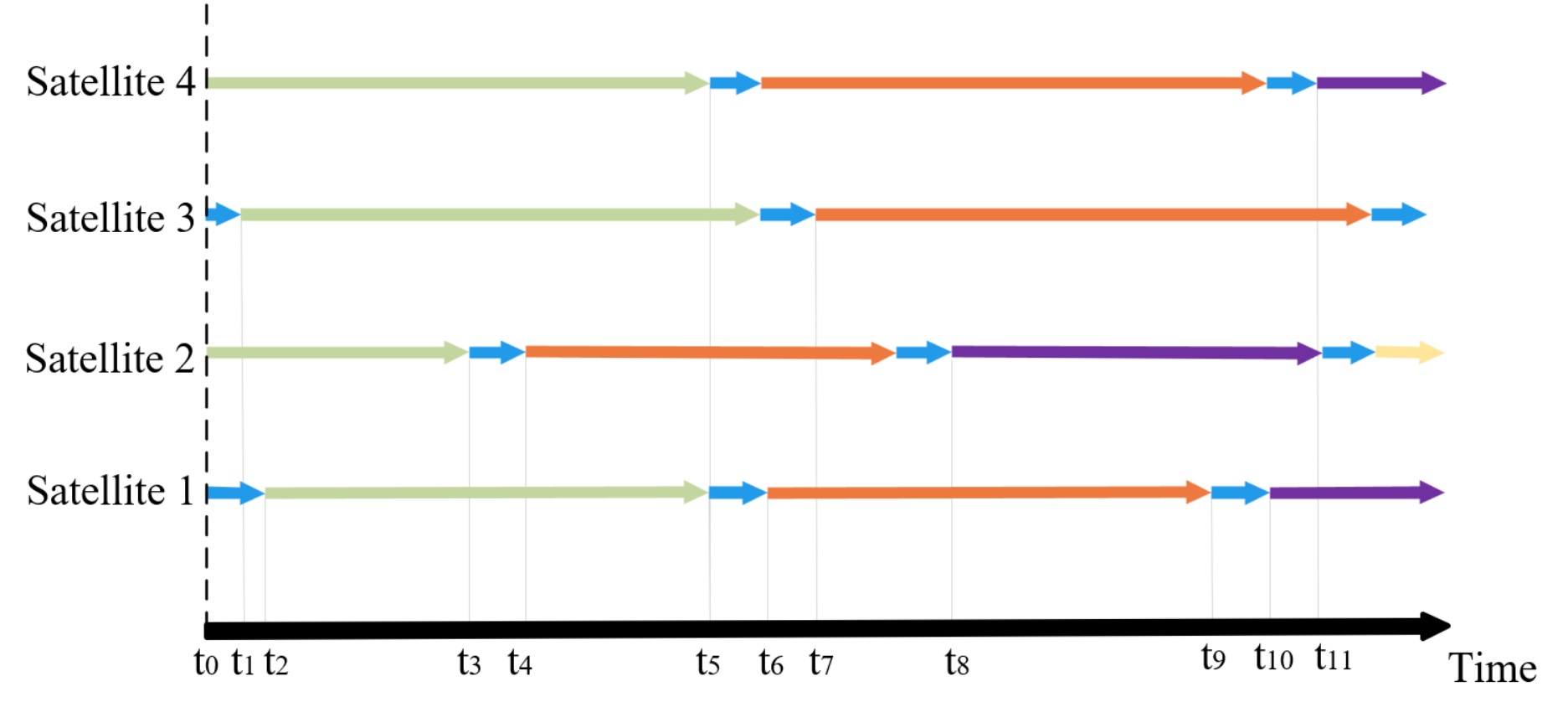}
\caption{The timing diagram of the asynchronous SFL in an illustrative example
with $N = 4, K = 1$.}
\label{fig:SA}
\end{figure}

In each round $i$ of the asynchronous algorithm, to make an update, the GS waits to receive the data from $K\leq N$ satellites sequentially. This set of satellites is denoted as $\mathcal{S}^i$. Note that the constellation follows a deterministic pattern, hence the smashed data of the satellites arrives at the GS in time order. More specifically, we define the $i$-th round as the time slot $[t_{Ki}, t_{K(i+1)})$, where $t_{Ki}$ denotes the instant when the $Ki$-th satellite becomes visible and hence the download communication can take place. 
In order to record the specific rounds in which each satellite $n\in \mathcal{N}$ contributes to the aggregation, an $N$-dimensional participation vector $\mathbf{r}_n$ is maintained by each satellite. For instance, if satellite $n$ participates in the $j$-th instance of round $i$, we denote the $j$-th element of $\mathbf{r}_n$ as $r_n(j) = i$. As shown in Fig.~\ref{fig:SA}, we have $N=4, K=1$, indicating that the GS updates after receiving results from $K=1$ satellite. In the first round, the GS collects the data from Satellite 1 and hence we have $\mathcal{S}^i=\{1\}$ and $r_1(1) = 1$.

We also define the initial global model as ${\bf{w}}^0=[{\bf{w}}_s^0,{\bf{w}}_g^0]$. Each satellite starts the local updates with the model ${\bf{w}}_{s,n}^{r_n(j),0}$, where the corresponding vector $\mathbf{r}_n$ is initialized to be zero. Once the split strategy is determined, the training process can be carried out. As a start, we pre-train our model and load the parameters into our model. We now present the procedure of each training round consisting of the following four steps, as illustrated in Fig.\ref{fig:plot_abstract}.

\textbf{Step 1: model download.} At each round $i$, once visible, each satellite $n\in \mathcal{S}^i$ downloads the current parameters ${\bf{w}}_s^i$ and $\mathbf{a}_s^i$ from the GS in time order and sets the local parameters as 
\begin{align}
{\bf{w}}_{s,n}^{r_n(j),0}={\bf{w}}_s^i,~~~\mathbf{a}_{s,n}^{r_n(j),0}=\mathbf{a}_s^i.
\end{align}


\textbf{Step 2: local model update and smashed data upload.} With the current local model ${\bf{w}}_{s,n}^{r_n(j),0}$ and a randomly selected mini-batch $\tilde{\mathcal{D}}_n \subset \mathcal{D}_n$, each satellite $n\in \mathcal{S}^i$ conducts $U$ iterative local updates. Specifically, for each iteration $u=0,\ldots, U-1$, satellite $n$ performs a feedforward pass up to the last layer of the auxiliary network $\mathbf{a}_{s,n}^{r_n(j),u}$ to compute the loss function. The local model and the auxiliary network are then updated through backpropagation, given as 
\begin{equation}
{\bf{w}}_{s,n}^{r_n(j), u+1}={\bf{w}}_{s,n}^{r_n(j),u}-\eta_i \bm{\nabla}_{{\bf{w}}} F_{s,n}({\bf{w}}_{s,n}^{r_n(j),u}, \mathbf{a}_{s,n}^{r_n(j),u}),
\end{equation}
\begin{equation}
\mathbf{a}_{s,n}^{r_n(j),u+1}=\mathbf{a}_{s,n}^{r_n(j),u}-\eta_i \bm{\nabla}_{{\bf{a}}} F_{s,n}({\bf{w}}_{s,n}^{r_n(j),u}, \mathbf{a}_{s,n}^{r_n(j),u}),
\end{equation}
where $\eta_i$ is the learning rate at round $i$ and $\bm{\nabla}F_{s,n}({\bf{w}}_{s,n}^{r_n(j)}, \mathbf{a}_{s,n}^{r_n(j)})$ is the derivative given a mini-batch, i.e., we have $\bm{\nabla}F_{s,n}({\bf{w}}_{s,n}^{r_n(j)}, \mathbf{a}_{s,n}^{r_n(j)})= \frac{1}{d_n}\sum_{x \in \tilde{\mathcal{D}}_n} l_n\left(x,{\bf{w}}_{s,n}^{r_n(j)}, \mathbf{a}_{s,n}^{r_n(j)} \right)$. 

To reduce storage and communication overhead, each satellite selectively preserves and transmits data from the most recent $u'$ local updates. In simpler terms, this data encompasses the smashed data set [$g_{{\bf{w}}_{s,n}^{r_n(j), U-u'+1}} (x),\ldots,g_{{\bf{w}}_{s,n}^{r_n(j), U}} (x)$], comprising the outputs of the satellite-side model, and their corresponding data labels. As soon as the satellite becomes visible, it is promptly transmitted to the GS.



\textbf{Step 3: GS-side model update.} While the satellite-side
models are being updated using the local loss functions, the GS also updates the GS-side model in parallel. After receiving the smashed data and their corresponding labels from the satellites in the set $ \mathcal{S}^{i-1}$, the GS performs $P$ model updates by following the rule
\begin{equation}
{\bf{w}}_{g}^{i,p+1}={\bf{w}}_{g}^{i,p}-\eta_t \bm{\nabla}_{{\bf{w}}} F_{s,n}(g_{{\bf{w}}_{s,n}^{r_n(j),u}} (x), {\bf{w}}_{g}^{i,p}). 
\end{equation}
where we have defined the model ${\bf{w}}_{g}^{i,0}={\bf{w}}_{g}^{i}$ and the derivative $F_{s,n}(g_{{\bf{w}}_{s,n}^{r_n(j),u}} (x), {\bf{w}}_{g}^{i,p})= \frac{1}
{d_n}\sum_{{\bf{x}} \in \tilde{\mathcal{D}}_{n}} {l\left( g_{{\bf{w}}_{s,n}^{r_n(j),u}} (x), {\bf{w}}_{g}^{i,p} \right)}$ for any $p=0,\cdots,P-1$.
As a result, the GS-side model is updated as 
\begin{align*}
    {\bf{w}}_{g}^{i+1}={\bf{w}}_{g}^{i,P}.
\end{align*}


\textbf{Step 4: global aggregation.} To perform an asynchronous aggregation, the GS waits until it receives the information from all the satellites in the set $\mathcal{S}^i$. Once visible, satellite $n$ uploads the updated satellite-side model ${\bf{w}}_{s,n}^{r_n(j),U}$ to the GS. 

Note that for each satellite $n$, the GS has updated $r_n\left(j\right) - r_n\left(j- 1\right)$ times before its visible period. In other words, the information returned by each satellite $n$ is computed at a stale value of the global parameter. The staleness $\tau_{n}^{r_n\left(j\right)}$ hence can be given as 
\begin{equation}
    \tau_{n}^{r_n\left(j\right)} \triangleq r_n\left(j\right) - r_n\left(j- 1\right).
\end{equation}
The staleness $\tau_{n}^{r_n\left(j\right)}$ can decelerate convergence and even intensify divergence of the global model significantly. Considering the inherent instability of the link between the satellites and the ground station, there is a potential risk of connection failure, which can increase data staleness. To address this issue and identify workers with high staleness, we introduce a threshold, denoted as $\tau_{max}$. At round $t$, if we have $\tau_{n}^{r_n\left(j\right)}> \tau_{max}$ for a satellite $n$, it refrains from transmitting over outdated data to the GS. Instead, it retrieves the latest global model from the GS once visible and commences a new round of computation. 


Hence, to penalize the impact, the GS aggregates the weighted satellite-side model according to
\begin{equation}
   {\bf{w}}_{s}^{i+1}=\frac{1}{K}\sum_{n\in\mathcal{S}^i}{p_n}{\bf{w}}_{s,n}^{r_n(j),U},
\end{equation}
where we have defined the penalty $p_n$ as 
\begin{equation}
   p_n=
    \begin{cases}
      \Big[1+\tau_{n}^{r_n\left(j\right)}\Big]^z, \text{if $\tau_{n}^{r_n\left(j\right)}> \tau_{max}$}\\
      0, ~~~~~~~~~~~~~~~~~~~\text{otherwise},
    \end{cases}\label{tau}
\end{equation}
with $z$ being some integer. 


The training processing repeats for $I$ rounds until the desired convergence criterion is
satisfied. This yields the global model ${\bf{w}}^I=[{\bf{w}}_s^I, {\bf{w}}_g^I ]$.


 \begin{figure}[t!]
\centering
\includegraphics[width=3in]{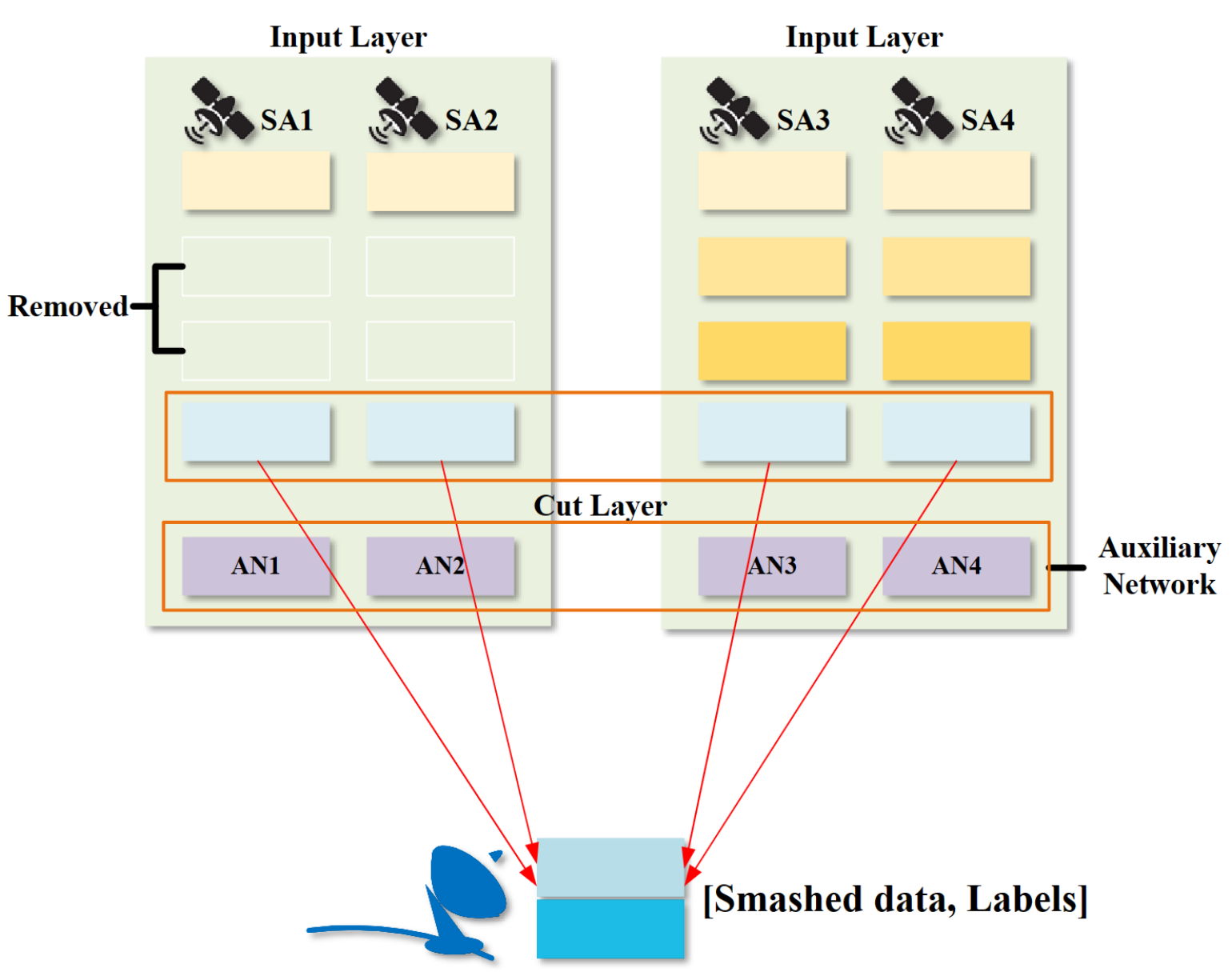}
\caption{The framework of personalized SFL with satellites having diverse computing capabilities.}
\label{fig:plot_abstract2}
\end{figure}

\subsection{Personalized Model split}
Given the varying computational capacities of the satellites alongside the scale of the large model, a crucial question emerges: how can we effectively partition the model to accommodate these differences? In this study, we specifically address a scenario where the satellite network consists of two types of satellites: those with strong computing capacities and those with weaker capabilities. When we partition the model, we allocate more layers to the former type, allowing them to handle more computations. 


It is important to note that this approach can be generalized to scenarios where each satellite possesses a unique computing capacity. By tailoring the model partitioning based on the specific capabilities of the satellites, we can optimize the performance and efficiency of the overall system.

In light of the spectrum of computational diversity, personalized model segmentation comes into play. To be specific, within this context, every individual satellite $n$ adopts a dedicated subsection $\mathbf{w}_{s,n}$ extracted from the overarching $\mathbf{w}_{s}$ model. Furthermore, to differentiate the satellites based on their varying computing capabilities, we classify the satellites into $M$ subsets denoted as $\mathcal{S}_1, \mathcal{S}_2, \cdots, \mathcal{S}_M$, with each subset representing a specific level of computing capability. The subsets are arranged in ascending order of computing capabilities, meaning that the first subset $g_1$ has the weakest computing capabilities among all the groups. 

For this scenario, we apply a novel strategy that lies in the following two aspects. 

\textbf{1. Keep the cut layer the same.} It is important to note that all the local models $\{\mathbf{w}_{s,n}\}_{n=1}^{N}$ have the same cut layer. The advantages of this approach are twofold. Firstly, it facilitates the connection between each satellite and the auxiliary network and ground stations. Secondly, this system empowers ground stations with the capacity to harness segmentation parameters extracted from all satellites for the purpose of model aggregation. This collaborative approach enhances the system's overall performance by pooling the strengths of multiple satellites, ultimately resulting in a higher level of model generalization and accuracy. For example, in Fig.~\ref{fig:plot_abstract2}, satellites 1 and 2 are from the subset $\mathcal{S}_1$, and satellites 3 and 4 are from the subset $\mathcal{S}_2$ with a stronger computing capability. Based on this, the satellites from the subset $\mathcal{S}_1$ compute fewer layers than the ones from the subset $\mathcal{S}_2$, which have more middle layers. This is achieved by applying the proposed splitting network layer alignment. 

 \begin{figure}[t!]
\centering
\includegraphics[width=3.2in]{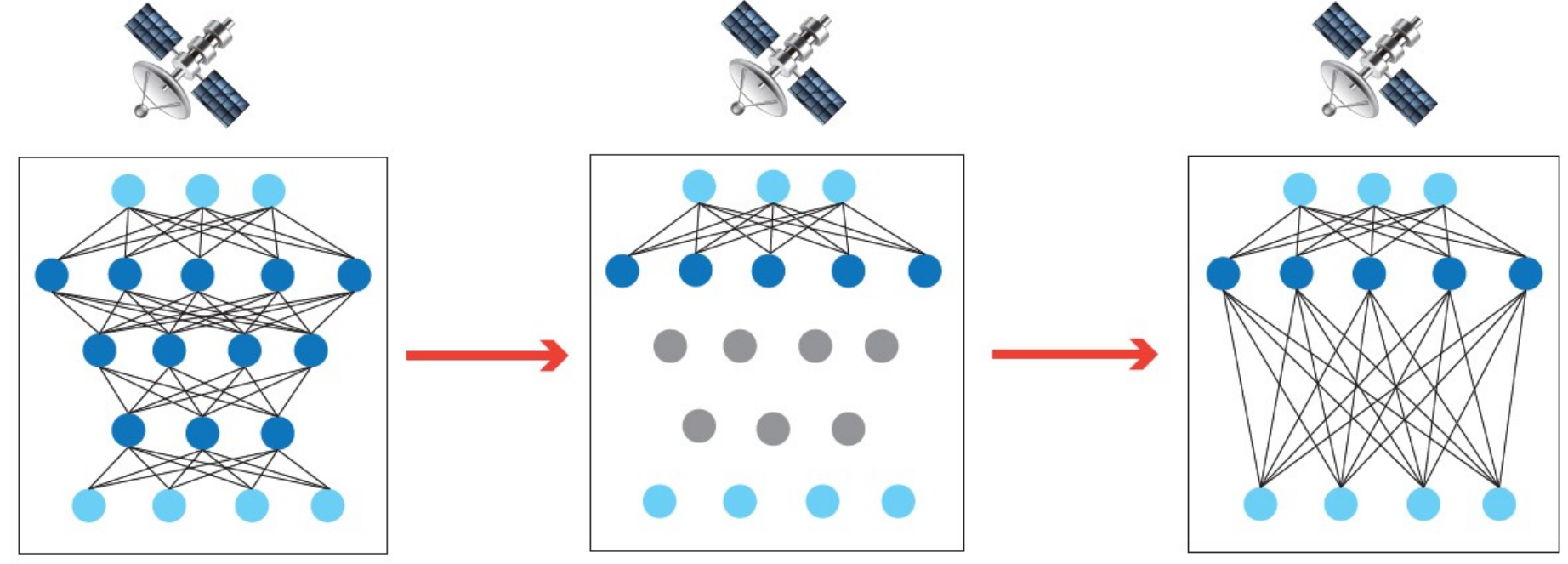}
\caption{Each satellite with weak computing capability has a personalized local mode by removing the middle layers. }
\label{fig:plot_abstract3}
\end{figure}

\textbf{2. Splitting network layer alignment.} To achieve this, for satellites with limited computing capabilities, we employ an adaptive approach to selectively remove intermediate layers, as depicted in Fig.~ \ref{fig:plot_abstract3}.

To elaborate, we maintain a consistent cut layer denoted as $\mathbf{w}_{s,n}^{c}$ across all satellites, referred to as the common layer. The individual layers unique to $\mathbf{w}_{s,n}$ are denoted as $\mathbf{w}_{s,n}^{up}$. This arrangement can be expressed as follows:
\begin{equation}
    \mathbf{w}_{s,n} =[\mathbf{w}_{s,n}^{up} , \mathbf{w}_{s,n}^{c}] .
\end{equation}
Subsequently, we categorize the satellites into distinct groups based on their $\mathbf{w}_{s,n}^{up}$ values, which we represent using the list $\mathcal{S}_1, \ldots, \mathcal{S}_M$. It's noteworthy that the initial three steps of the personalized model split procedure for each training round remain identical to the standard model split.

Now we focus on the global aggregation step. Before each aggregation, satellite $n$ which we assumed belongs to $S_t$ uploads its updated satellite-side model $\mathbf{w}_{s,n}$ to the GS. The GS should recognize the individual layer $\mathbf{w}_{s,n}^{up}$ and the common layer $\mathbf{w}_{s,n}^{c}$ of it and aggregate them respectively.
As for asynchronous aggregation, we keep the same operation of staleness as identical model split. And the model aggregation details are shown as following 
\begin{equation}
   {\bf{w}}_{s}^{i+1,up}=\sum_{n\in\mathcal{S}^i,n\in S_t}p_{n}{\bf{w}}_{s,n}^{r_n(j),U,up} .
\end{equation}
\begin{equation}
    {\bf{w}}_{s}^{i+1,c}=\sum_{n\in\mathcal{S}^i}p_{n}{\bf{w}}_{s,n}^{r_n(j),U,c} .
\end{equation}
\begin{equation}
   {\bf{w}}_{s}^{i+1}={\bf{w}}_{s}^{i+1,up} + {\bf{w}}_{s}^{i+1,c} .
\end{equation}
We give pseudo code to introduce the whole procedure of this algorithm.
For this case, in the presence of a straggler, can we choose only one within each group? or if we have one in each group, can we drop the straggler one?

\begin{algorithm}[t!]
\caption{SFL\_over\_SA Algorithm}
\label{a1}
\begin{algorithmic}
\STATE Initializes $K$,  $\mathbf{a}_{s}^{0}$, $\mathbf{W}^0= [\mathbf{W}_s^0,\mathbf{W}_g^0]$, $N$, $T$, $j = [0,0,\ldots, 0]$
\STATE \FOR{satellite $n \in N$}
\STATE Distribute initial model layers to satellite $n$ according to $w_{s,n}^{up}, w_{s,n}^{c}$ 
\ENDFOR
\STATE \FOR{$i=1,\dots,T$}
\STATE Focus on current satellite $n$
\STATE \textbf{Do Local Train of $n$:}
\begin{ALC@g}
\STATE \FOR{each epoch round $u$}
\STATE Update $w_{s,n}^{r_n(j[n]),u}$ and $a_{s,n}^{r_n(j[n]),u}$ via backward gradient 
\ENDFOR
\STATE Put $n$ into $S^i$
\STATE Get model parameters, smashed list of $n$
\end{ALC@g} 
\STATE \IF{$n$ is visible}
\STATE \FOR{(smashed data, smashed labels) in smashed list}
\STATE Train server with $W^0$, smashed data, smashed labels
\ENDFOR
\STATE Set $r_n(j[n])=i$
\STATE Set $j[n] = j[n] + 1$
\ENDIF
\STATE \IF{$|S^i| \geq K$}
\STATE Do model aggregate using model parameters of satellites from ${S}^i$ and model parameters aggregated last time
\STATE Clear up ${S}^i$
\ENDIF
\STATE Get loss and accuracy of train and test sets
\STATE Send model parameters after aggregation back to satellite $n$
\ENDFOR

\end{algorithmic}
\end{algorithm}

\section{Experiments}
In this section, we evaluate the proposed SFL-LEO algorithm on CIFAR-10. We compare with three baseline algorithms, i.e., vanilla FL, corresponding to the case with SL scheme as well as centre approach.

\subsection{Experimetal Setup}
\textbf{Data Distribution.} We distribute the training set of each dataset to the clients for training, and utilized the original test set of each dataset to evaluate the performance of the global model. We consider a system with $N =20$ satellites. Hence, each client has 2500 data samples for CIFAR-10. Here, we
consider two different data distribution setups, IID and non-IID setups. In an IID setup, data samples from each class is equally distributed across all $N$ satellites in the system. Hence, each satellite has all 10 classes in its local dataset. In a non-IID setup, the training set is first divided into 5000 shards (10 data samples in each shard for CIFAR-10). Then we allocate 250 shards to each client to model the non-IID scenario.

\textbf{Model splitting and the auxiliary network.} To comprehensively simulate various tasks on satellites, we employed two distinct neural network models of different sizes in our experiments.

First, we defined an VGG16Net \cite{DBLP:journals/corr/SimonyanZ14a}. Concretely, we list out the parameters of the first 4 convolutional layers with incremental order of layers by 5-tuple form which means (in channels, out channels, kernel size, stride, padding) respectively: (3, 64, 3, 1, 1), (64, 64, 3, 1, 1), (64, 128, 3, 1, 1), (128, 128, 3, 1, 1). Two schemes of splitting network were come up by us. The first one is cutting the first 2 layers of network for all satellites and the ground station obtains the remaining layers. The other is cutting 2 layers which are the first and fourth layers of VGG16Net for satellites with low computation ability and the first 4 layers for other satellites when the ground station obtains the remaining layers. Especially, the 5-tuple for 2-layer satellites are (3, 128, 3, 1, 1) and (128, 128, 3, 1, 1) which we can see that we adjust the in the channel from 64 to 128 for the correct connect between 1st and 4th layer after removing 2nd and 3rd layer guaranteeing all the satellites keep the same 5-tuple for their cut layer. Furthermore, we put an auxiliary network which is simply a fully-connected layer between the cut layer of satellites and the ground station to help satellites to do local training without connecting to the ground station.

Secondly, we designed a lightweight neural network, which we refer to as VGG16-mini, based on the architecture of VGG16Net. In this design, we have downsized each layer of VGG16Net by a factor of 8. Specifically, the first layer of the network employs 8 convolutional kernels instead of the original 64. Corresponding adjustments have been made to other segments of the network, as well as to the partitioning strategy and auxiliary networks.

\textbf{Implementation Details.} We consider a dynamic attenuation learning rate $\eta$ from $0.1$ to $0.001$ for ground station. We distinguished the learning rate between satellite and ground station and set satellite's only $1/40$ of ground stations. 
We choose $z=-4$ for $p_n$ to handle the staleness of the model.

To simulate the computational capabilities of the ground station, we utilized a server equipped with a NVIDIA GeForce RTX 4090 GPU. For the satellites' simulations, we utilized the server's CPU (16 cores, 2.5GHz). Referring to the data from the satellite ground station \cite{starlinkbandwidth}, the bandwidth of the satellite-ground link was configured to be 2 Gbps.

\textbf{Comparison schemes.} We compare our proposed learning scheme with the following three learning schemes, in order to demonstrate the method's capability of supporting satellite local updates while minimizing the impact on accuracy.
\begin{itemize}
\item \textbf{SL} refers to a classical SL scheme where satellites only train when connected to the GS. In our experiments, satellites were connected to the ground station for about 5\% of the time.

\item \textbf{FL} refers to a classical FL scheme where every satellite keep the whole model. Considering the limit computing resources onborad, this scheme is impractical in reality. In the experiments, we assumed that the resources on the satellites were sufficient for computation.

\item \textbf{Centre} refers to the approach where all data is transferred to the GS, and a complete network is trained at the GS. This approach is expected to yield the highest accuracy. However, this scheme is not feasible in the real world due to the data privacy concerns.
\end{itemize}

\subsection{Experiment Results}

\begin{figure}[t!]
\centering
\includegraphics[width=3in]{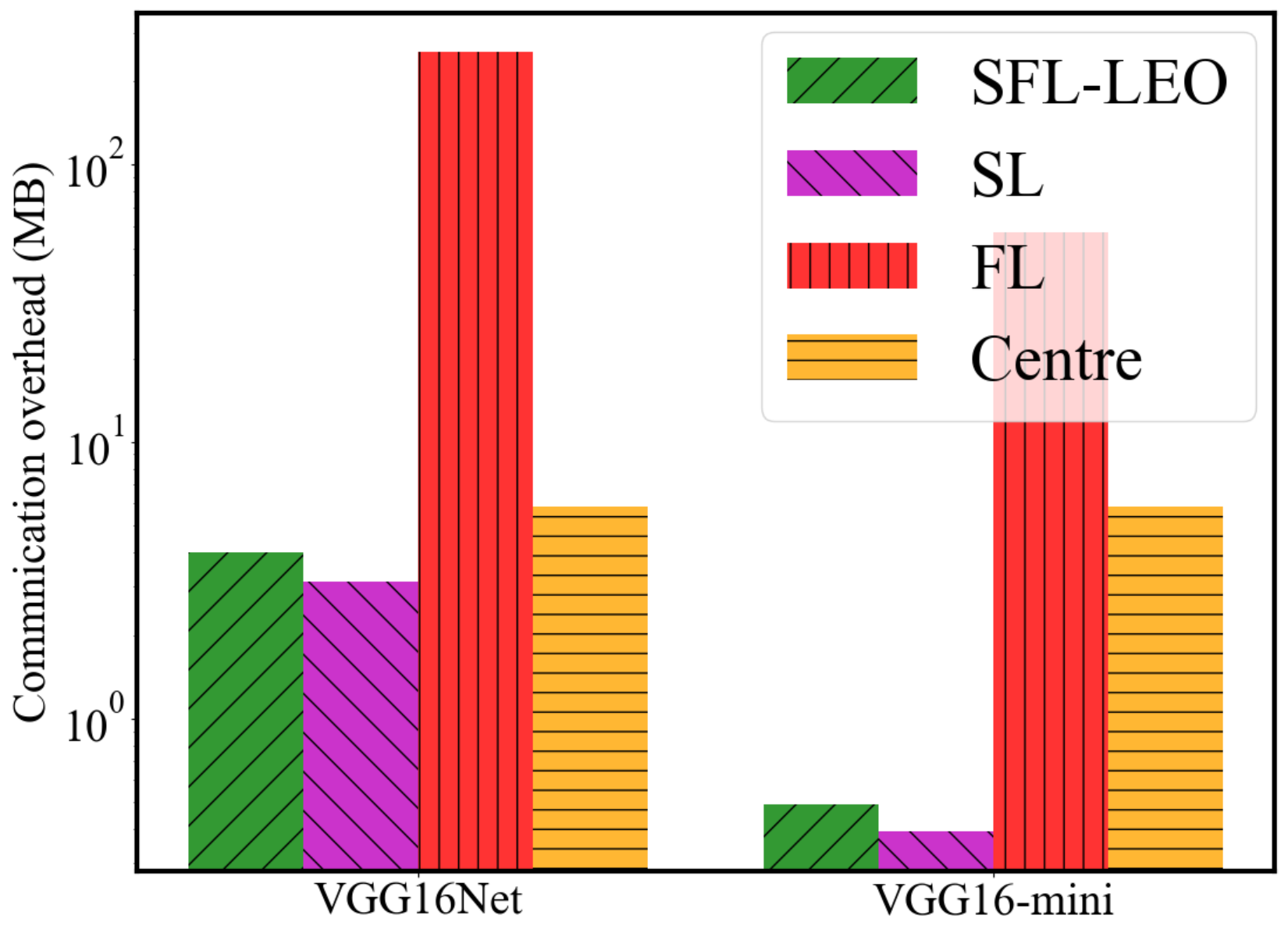}
\caption{Communication overhead for each satellite each cycle.}
\label{fig:communication_overhead}
\end{figure}

\noindent\textbf{Communication overhead:} First, we conducted a comparison of the algorithm's communication overhead. As illustrated in Fig. \ref{fig:communication_overhead}, our approach significantly reduces the volume of transmitted data compared to the centre scheme, which necessitate the transmission of all raw data. In comparison to the SL method, our approach entails a similar information transmission size. Given the constraints of satellite-to-ground link bandwidth, we consider this to be an appropriate optimization. 

\begin{figure}[t!]
\centering
\includegraphics[width=3in]{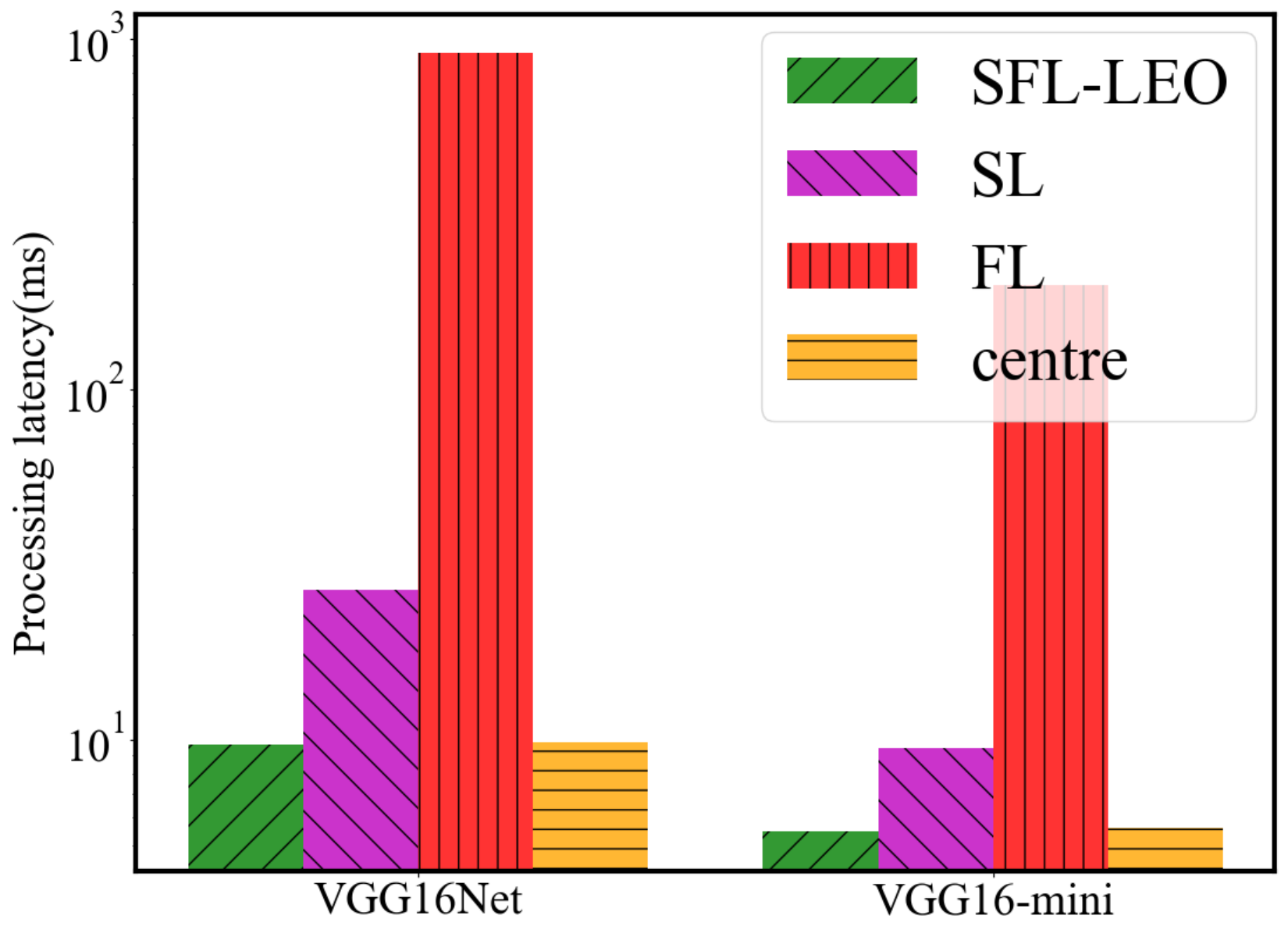}
\caption{Latency pf processing new data.}
\label{fig:speed}
\end{figure}

\noindent\textbf{Time of processing data}
Furthermore, we conducted a comparison of latency. In this context, we considered the delay from the moment a satellite connects to when it completes model updates. As depicted in Fig. \ref{fig:speed}, we also compared scenarios involving two different models. It's evident that our approach significantly reduces processing latency for average 65 $\times$ compared to the FL approach. This reduction occurs because the SFL algorithm not only minimize the transmitted data volume but also reduce the size of the model processed on the satellite. Compared to the SL approach, SFL gains an reduction in processing latency for about $56\%$ because it performs local updates during disconnected periods. In comparison to the center-based scheme, SFL's time expenditure is similar. This similarity is attributed to the fact that both schemes rely on ground-based servers for primary computations, and since the data transmission volume is minimal, this outcome is reasonable.

\begin{figure}[t!]
\centering
\subfloat[IID Data]{\includegraphics[width=0.5\linewidth]{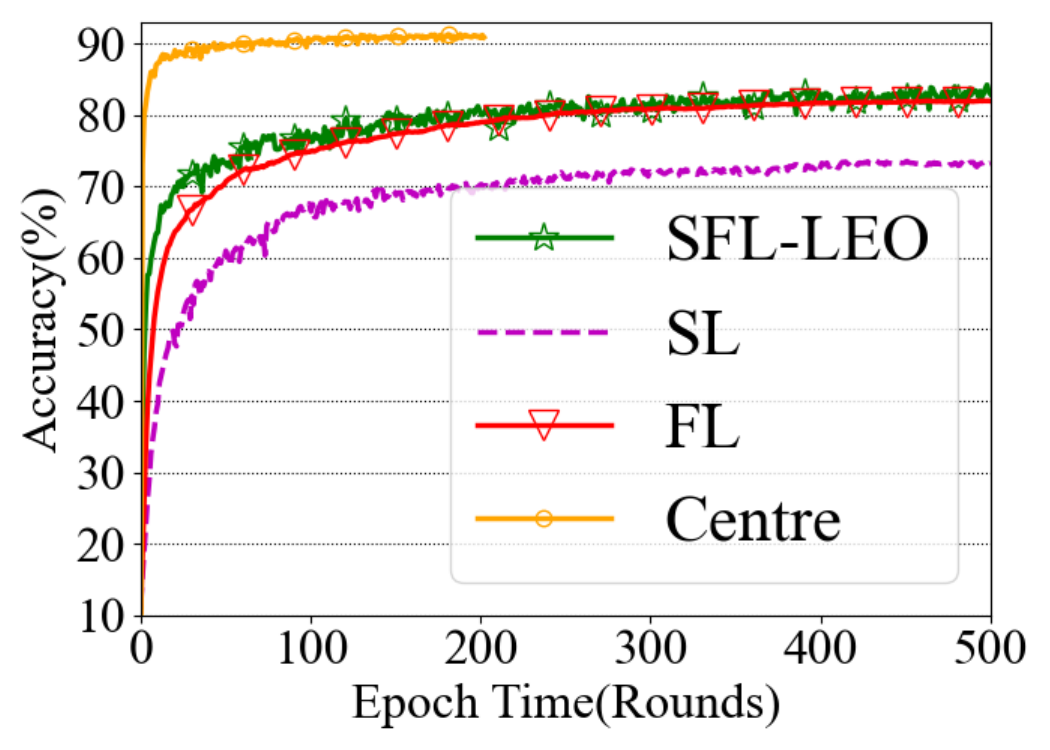}\label{fig:iid_method}}
\subfloat[Non-IID Data]{\includegraphics[width=0.5\linewidth]{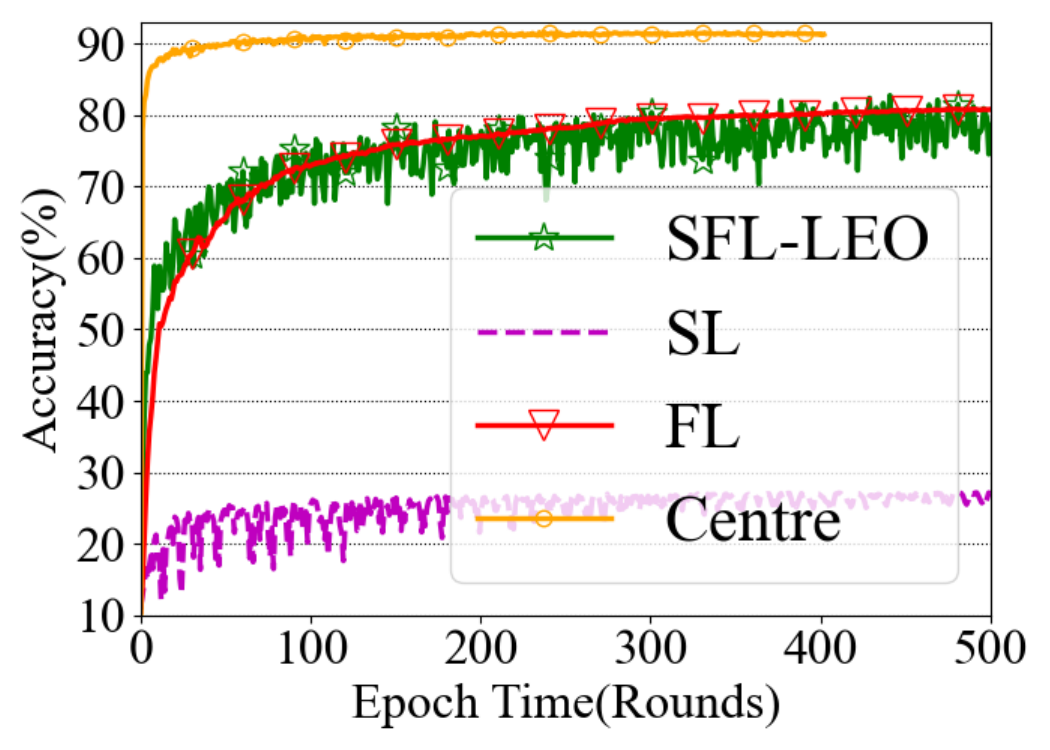}\label{fig:niid_method}}
\caption{Test accuracy with different learning schemes and data during training.}
\label{fig:methods}
\end{figure}

\noindent \textbf{Test accuracy among different learning schemes:} As shown in Fig. \ref{fig:methods}, we conduct a comparative analysis of different learning schemes' test accuracy under varying data conditions. It can be observed that the proposed SFL-LEO scheme demonstrates comparable or even superior performance compared to existing approaches. More specifically, as illustrated in Fig. \ref{fig:iid_method}, when dealing with data following an IID distribution, the centre scheme, as an ideal scheme, achieves the accuracy of 91.2 \% with the fastest convergence speed. The three distributed training schemes, including the proposed SFL-LEO scheme, the SL scheme, and the FL scheme, exhibit similar convergence speeds. Among these, SFL-LEO achieve an accuracy of 84.4\%, which is higher than taht of other 2 schemes. The FL scheme attains an accuracy of 83.2\%, and the SL scheme attains an accuracy of 73.2\%.

It is not surprising that the SL scheme exhibits the poorest performance. Under this scheme, computation only takes place when a connection is established, resulting in significantly less data utilization and fewer training iterations compared to the other schemes. An intriguing observation is the performance of the FL method, which, as the most demanding approach for each satellite, yields the poorer accuracy compared to the purposed SFL-LEO scheme and centre scheme. We believe that this phenomenon can be attributed primarily to the relatively weaker aggregation capability of the FL mechanism in handling the training outcomes from various satellites.



Fig. \ref{fig:niid_method} describes the accuracy when using the non-IID data. Due to the influence of data distribution disparities, both SL and SFL experience fluctuations and a loss in accuracy. In this situation, the SL only get an accuracy of less than 30\% as its training rounds are too few. However, it is noticeable that the accuracy degradation and fluctuations in SFL are significantly less pronounced than in SL.
This effect is attributed to the more efficient aggregation strategy employed in the SFL scheme, which effectively leverages data from different satellites. 
\begin{figure}[t!]
\centering
\subfloat[SFL (ideal)]{\includegraphics[width=\linewidth]{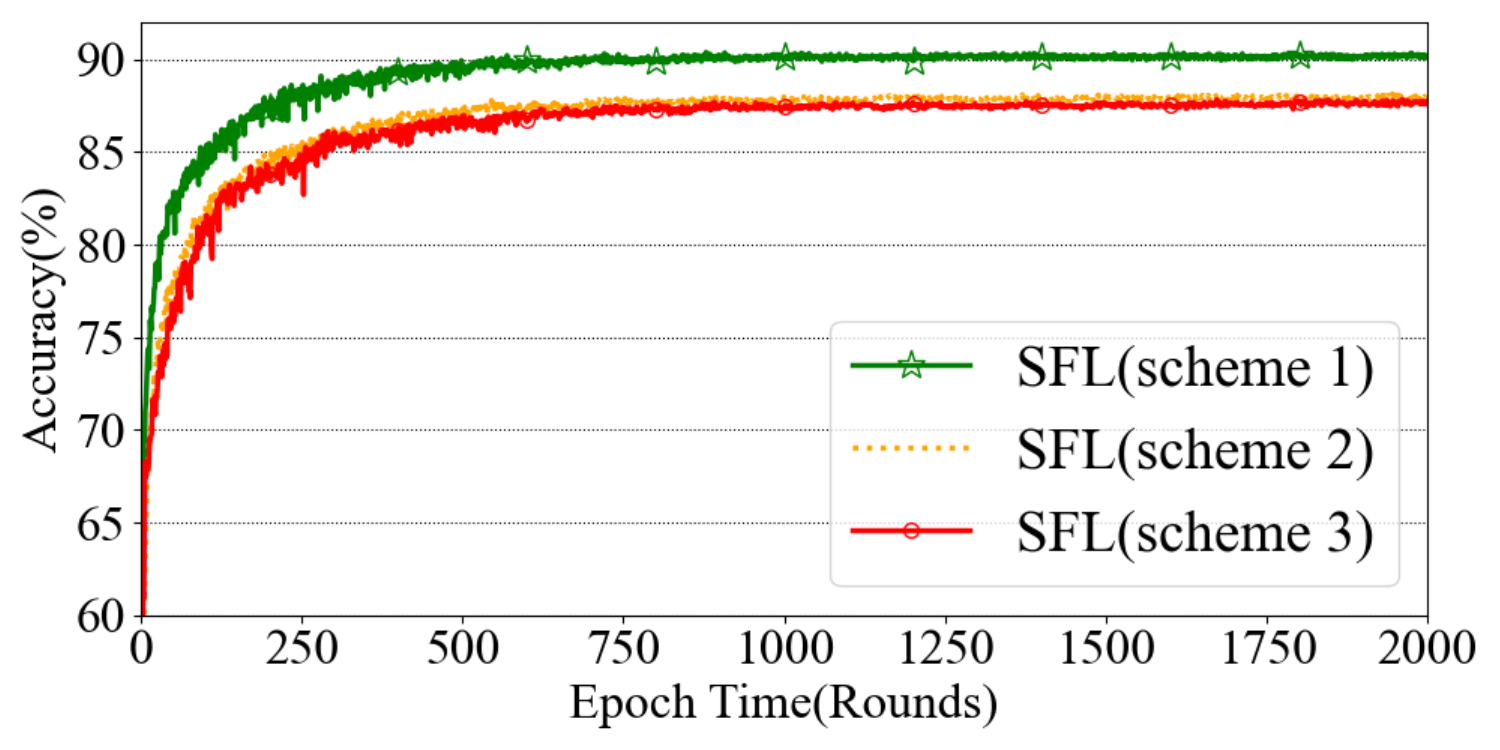} \label{fig:SFLsplit}}

\subfloat[SL (ideal)]{\includegraphics[width=\linewidth]{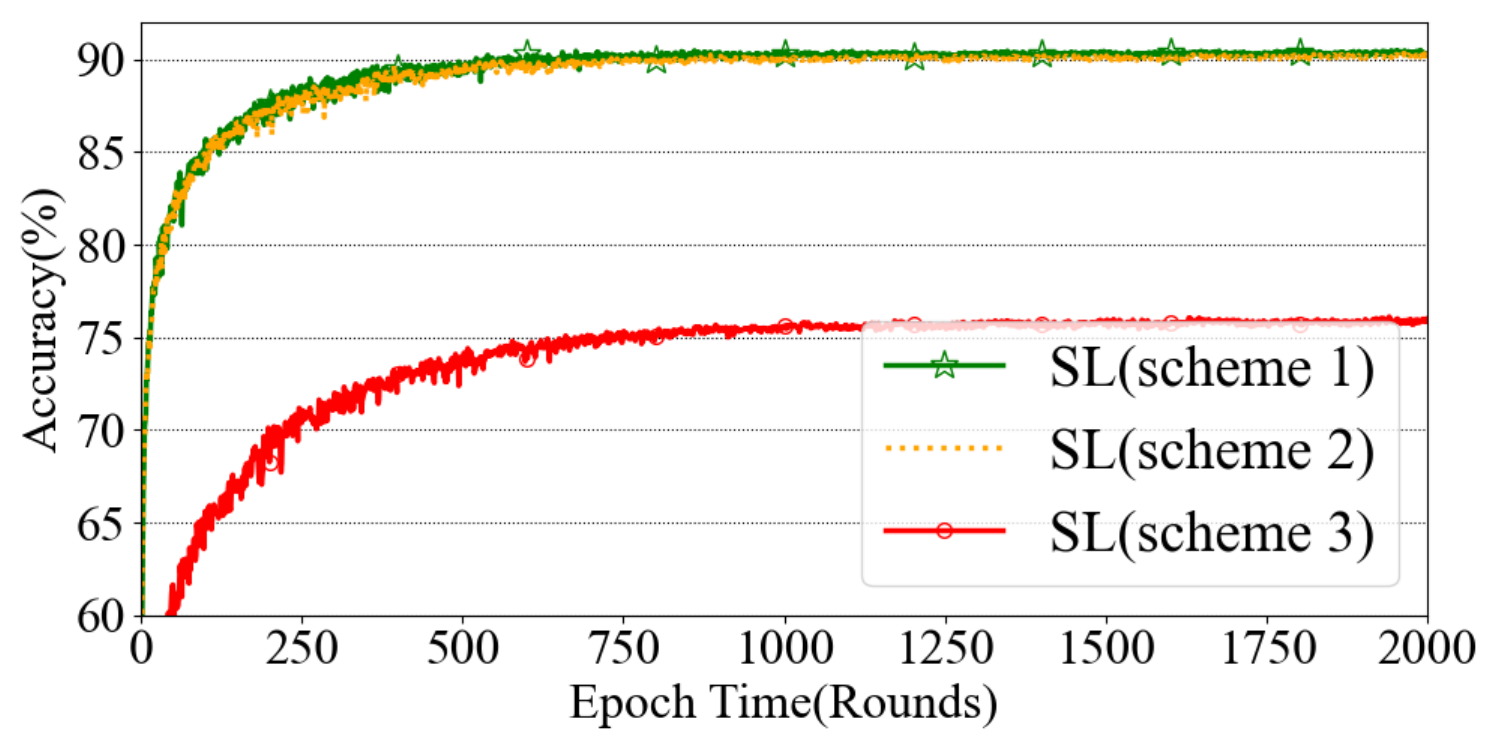} \label{fig:SLsplit}} 
\caption{Influence of different splitting networks.}
\label{fig:split}
\end{figure}

Furthermore, we conducted experiments to assess the impact of different network split schemes on overall performance. This includes scenarios where different split network sizes are used for satellites employing the same network structure, as well as scenarios where different splitting network structure are present within the network. Given that the center and FL schemes do not involve network splitting, we select the SL scheme as the comparative benchmark. In order to facilitate a fairer comparison, we made adjustments to both the SFL and SL schemes to ensure that the data volume and training iterations were comparable. Specifically, we assumed that in the SL scheme, satellites remained connected to the ground station throughout, enabling continuous training. As a result, the amount of data transmitted in the SFL scheme was accordingly increased. These modified schemes are referred to as SFL (ideal) and SL (ideal), respectively.

We examine two different split size choices under the same network structure scenario, along with the situation where two distinct split network architecture coexist. The size of each split networks are outlined in Table \ref{table:param}.

\begin{table}[t!]
\caption{Detail of different split schemes}
\begin{center}
\begin{tabular}{|c|c|c|c|}
\hline
    &scheme 1 & scheme 2 &scheme 3\\
    \hline
    Split stragety & \multicolumn{2}{c|}{same networks}&different networks  \\
    \hline
     GS Param& \multicolumn{3}{c|}{$5.4*10^8$} \\
     \hline
     \multirow{2}{*}{Sat Param}&\multirow{2}{*}{$1.6*10^5$}&\multirow{2}{*}{$1.0*10^6$}&$6.0*10^5$\\
     \cline{4-4}
     \multirow{2}{*}{}&\multirow{2}{*}{}&\multirow{2}{*}{}&$1.0*10^6$\\
     \hline
\end{tabular} \label{table:param}
\end{center}
\end{table} 

Fig.\ref{fig:SFLsplit} and Fig.\ref{fig:SLsplit} depict the performance of SFL and SL approaches under different split schemes respectively. The varying performances between scheme 1 and scheme 2 highlight the impact of different split network sizes. Notably, a trend emerges: as the number of parameters onboard the satellite diminishes, the corresponding accuracy increases. This relationship is logically expected. Specifically, when the onboard parameters reach zero, the SFL-LEO approach transitions into a mode akin to centre scheme. Conversely, as the onboard parameters align with those on the ground, the approach aligns more closely with an FL scheme. 

On the other hand, the comparison between scheme 2 and scheme 3 illustrates the impact of coexisting multiple splitting network structures within the network. The findings reveal that the accuracy of the SL approach undergoes significant degradation when multiple split network structures coexist. In contrast, the SFL-LEO demonstrates performance that's on par with the the same network structure scenario. This attests to our method's capacity to adeptly accommodate diverse split structures within the network, effectively addressing the challenges arising from heterogeneous computational capabilities.
\begin{figure}[t!]
\centering
\includegraphics[width=1\linewidth]{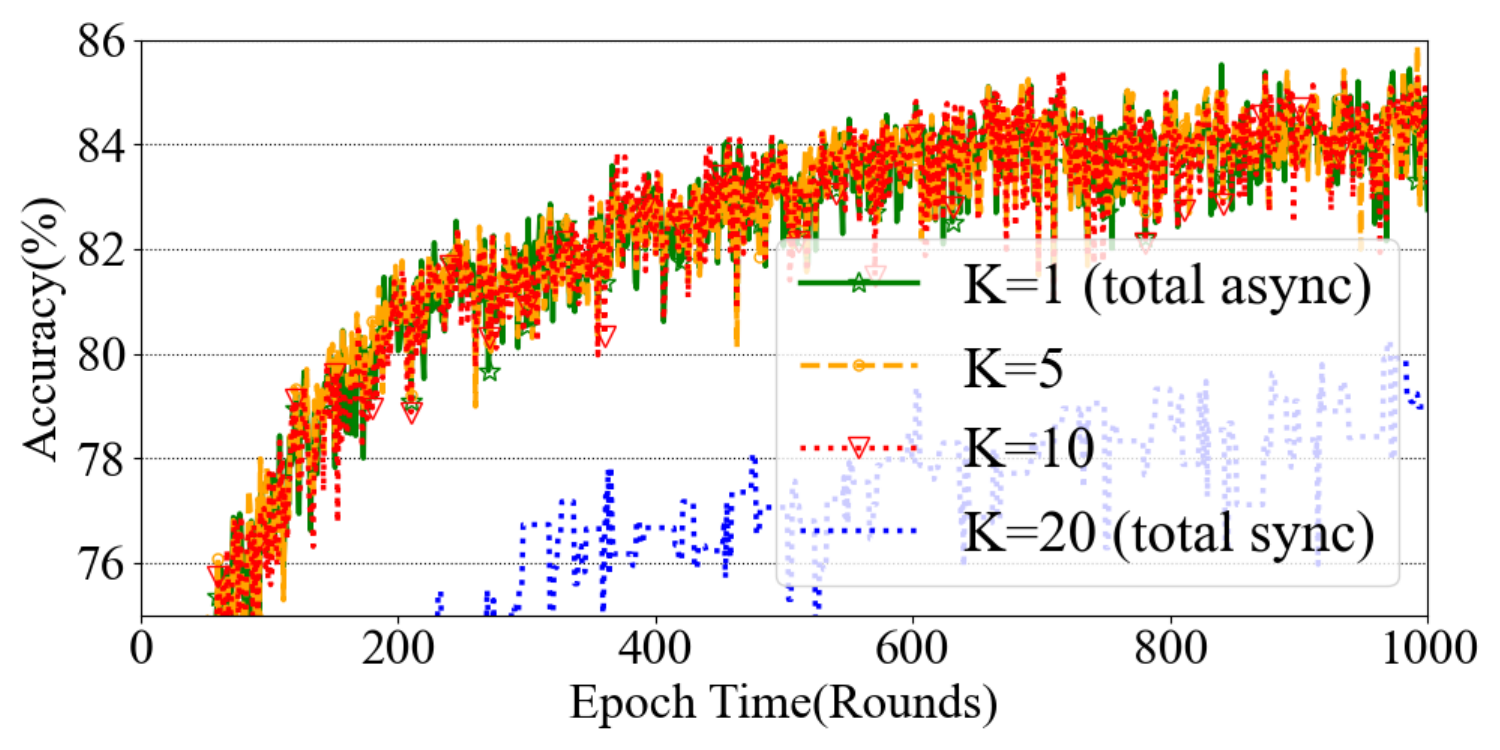}
\caption{Influence of asynchronous update.}
\label{fig:asyn}
\end{figure}


We also examined the impact of asynchronous updates on model performance. As depicted in Fig. \ref{fig:asyn}, when the model is completely asynchronous (i.e., $K=1$), it exhibits the best performance. However, as $K$ increases, indicating higher synchronization among models, the performance gradually diminishes. This phenomenon can be attributed to the fact that asynchrony enhances the richness of information within the model. Conversely, under complete synchronization situation, the aggregation model fails to utilize staleness, leading to the observed decline in performance.

\begin{figure}[t!]
\centering
\subfloat[IID data]{\includegraphics[width=0.5\linewidth]{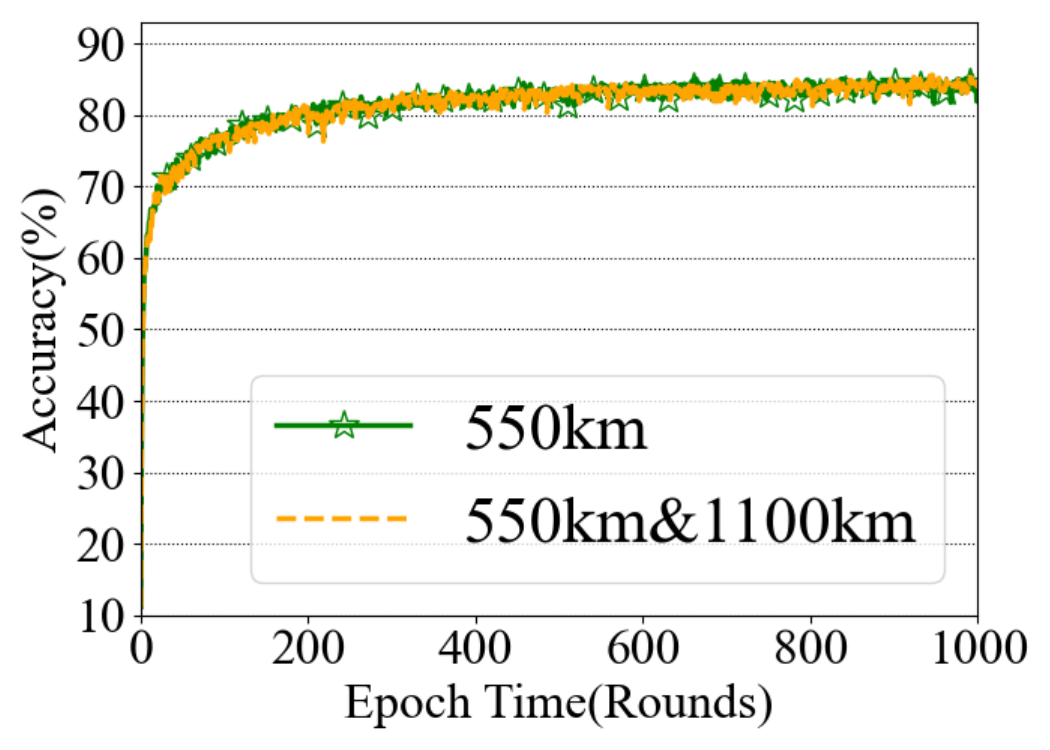} \label{fig:orbitiid}}
\subfloat[NonIID data]{\includegraphics[width=0.5\linewidth]{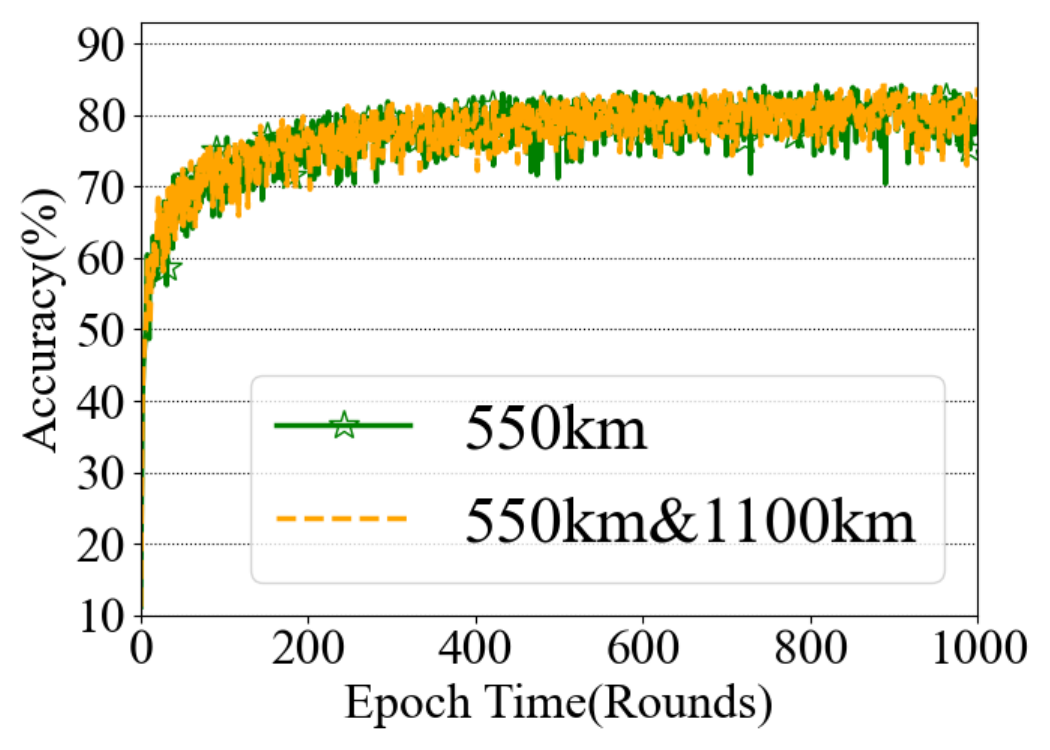} \label{fig:orbitniid}} 
\caption{Influence of different orbit.}
\label{fig:differnet_orbit_result}
\end{figure}
 
Taking into account the possibility of satellites at different orbital altitudes within the constellation, we conducted further experiments to validate the algorithm's effectiveness in such scenarios. Fig. \ref{fig:differnet_orbit_result} presents a comparison of the accuracy achieved by the SFL algorithm in two scenarios: one where all satellites are at an altitude of 550 km and another where satellites are distributed between altitudes of 550 km and 1100 km. The results indicate that, irrespective of whether the scenario follows an IID or Non-IID distribution, the algorithm consistently delivers similar performance, regardless of variations in satellite orbit altitudes. This observation suggests the algorithm's adaptability to varying satellite altitudes. This can be partly attributed to the SFL algorithm's utilization of local updates, which helps counteract the impact of varying connection time disparities arising from different altitudes on each satellite's data retrieval procedures. This adaptability underscores the algorithm's robustness across diverse satellite altitudes.

\begin{figure}[t!]
\centering
\subfloat[IID data]{\includegraphics[width=0.5\linewidth]{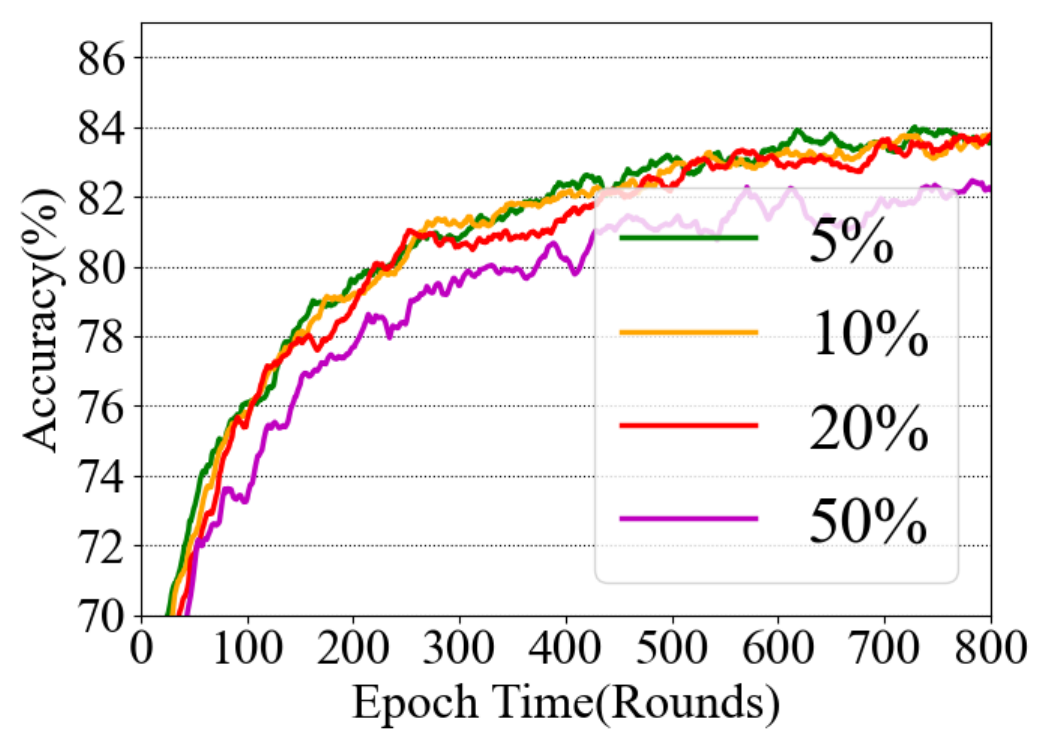} \label{fig:lossiid}}
\subfloat[NonIID data]{\includegraphics[width=0.5\linewidth]{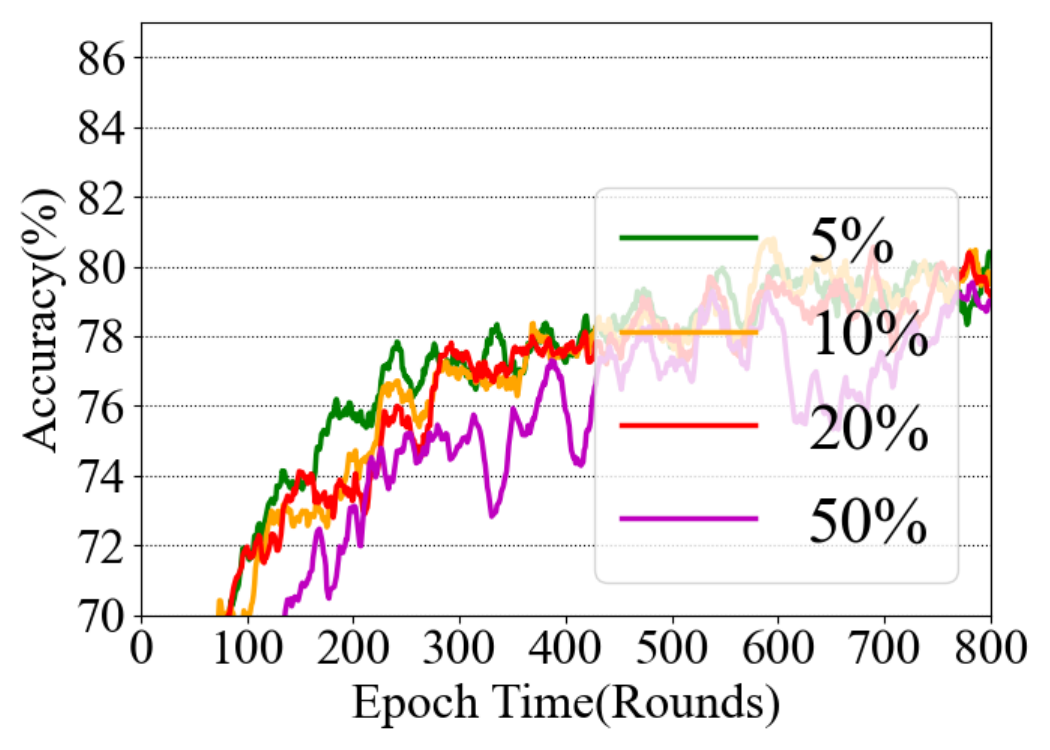} \label{fig:lossniid}} 
\caption{Performance under different satellite loss probability.}
\label{fig:differnet_loss}
\end{figure}

In consideration of potential disruptions in satellite-ground communication links, even during satellite visibility periods, interruptions or complete failures in establishing ground links can transpire. In an extreme scenario, external interferences, such as adverse weather conditions, can preclude the establishment of satellite-to-ground links, rendering satellite data inaccessible for processing at ground stations. To address this concern, a series of experiments were conducted. Fig. \ref{fig:differnet_loss} illustrates the algorithm's performance under varying probabilities of satellite loss. It is evident that as the rate of satellite disconnection increases, there is a gradual reduction in algorithm accuracy. Remarkably, even at a disconnection rate of 25\%, the impact on performance remains relatively modest. This observation underscores the robustness of our approach in the face of unstable satellite-ground communication links.

\section{Related Work}
Recently, the LEO satellite constellation construction has drawn widespread attentions from both industry and academia because it can provide worldwide Internet service to end users. Meanwhile, a large amount of data such as remote sensing image and baseband information collected by LEO satellites challenges the delivery from satellites to the ground due to the limited and fluctuated bandwidth between them. To overcome this challenge, available research can be divided into two categories.

First, there exists a large body of research work on improving the downlink performance, which can support high-bandwidth transmission \cite{carvalho2019optimizing,castaing2014scheduling}. However, these work still suffers from the bandwidth bottleneck and thus results in long transmission delay. Meanwhile, the authors of \cite{10.1145/3452296.3472932} formulate the link scheduling problem based on real collected weather data as a NP-hard problem, which is addressed by the proposed (approximate) greedy algorithm.   


Secondly, the abundant computation resources hidden in the huge satellite constellation provide a feasible solution for data processing at LEO satellites and reduce the amount of data transmitted from satellites to the ground station, which also contributes to the decrease of the overall delay \cite{razmi2022ground,matthiesen2023federated,chen2022satellite,10021101,lin2023split,lin2023efficient,lyu2023optimal}. For example, the authors of \cite{matthiesen2023federated} made a comprehensive analysis and comparison between several research work related to FL algorithms in satellite contexts. Meanwhile, it also investigates the design of FL scheme when taking Inter-Satellite Links (ISLs) and satellite visibility into consideration. However, above research work ignores the computation-heterogeneous problem among the huge satellite constellation, which still remains far from the real scenarios of satellite networks.  



To address the challenge of facilitating machine learning training on performance-constrained devices, such as satellites, and to enhance training efficiency, SFL and Parallel Split Learning (PSL) have been introduced. These approaches aim to parallelize the training of client-side models, enabling individual clients to simultaneously train sub-models \cite{oh2022locfedmix,lin2023efficient,chen2021communication,mu2023communication,hong2022efficient}. These research efforts focus on the ground scenarios where the static edge devices connect with the cloud and end terminals simultaneously. However, they are unsuitable for the LEO satellite networks coupled with high mobility and computation-heterogeneity. For instance, the short-duration connectivity between satellites and the ground station challenges the training model design.



Motivated by the recent research of combining the SF scheme with FL scheme \cite{thapa2022splitfed,HSFL2023}, we demonstrate how to skillfully integrate the FL scheme with the SL scheme and propose a hybrid SL and FL co-design for data processing, to overcome the challenges unique in LEO satellite networks. This work represents the first research efforts to achieving high-efficient computation that accommodates for the high dynamics and computation-heterogeneity in the LEO satellite-ground network topology.    


\section{Discussion and limitations}
SFL-LEO is a hybrid split-federated design for data processing in the satellite-ground network topology by intelligently combining the SL and FL schemes. As several advantages such as the decrease of the amount of transmitted data and executing delay can be brought by SFL-LEO, there still exist several limitations with SFL-LEO, which is described as below.   

\textbf{Implementation of SFL-LEO}: Due to the experiment constraints, we cannot deploy SFL-LEO on real LEO satellites. Instead, we have implemented SFL-LEO based on the built prototype and conducted extensive experiments which are driven by real Starlink traces and bandwidths, in order to simulate the real LEO satellite networks. Meanwhile, we have also simulated the computation capability of LEO satellites and ground station with highly similar hardware.

\textbf{Energy consumption}: SFL-LEO allows LEO satellites training locally when they disconnect with the ground station, which introduces extra energy consumption for LEO satellites. One may wonder whether this extra energy consumption is affordable for resource-constrained LEO satellites. Actually, this energy overhead is acceptable for two reasons. First, the split learning strategy is introduced to cut the overall model, where the computation overhead brought by training at satellites only occupies 1\% or even less of the overall model training overhead. Hence, this extra computation overhead in SFL-LEO brings a little energy consumption. In our experiments, the extra energy consumption caused by local update is around 50W. From another perspective, LEO satellites can provide up to several KW power consumption by the solar cell at LEO satellites. Consequently, the extra power consumption only occupies a small fraction of the power that LEO satellites can provide.  

\textbf{Safety problem}: SFL-LEO is a distributed learning framework~\cite{zhang2025lcfed,lin2024hierarchical,hu2024accelerating,lin2023split} targeting at solving the data processing problem in the satellite-ground topology. In fact, SFL-LEO can provide high security guarantee for data delivery between satellites and the ground station~\cite{lin2023efficient}, which is another important metric in satellite networks due to easy eavesdropping and attacks in open satellite-ground environments. We leave the safety improvement and measurement for future work because it is beyond the scope of this paper. 

\textbf{SFL-LEO's performance}: In the experiment part, we have made a comprehensive and in-depth performance comparison in terms of the amount of transmitted data, delay, training accuracy with the conventional SL and FL schemes, and the centre scheme. Based on these results, we can find that SFL-LEO is not a satisfactory design in all aspects. For example, compared to the centre scheme, SFL-LEO can reduce the amount of data transmitted from LEO satellites to the ground, yet with a lower training accuracy. If taking a detailed observation, SFL-LEO can considerably reduce the transmitted data which is a critical bottleneck in satellite-ground communications, while sacrificing a little training accuracy. For future work, we will make an attempt to increase the training accuracy with the most ideal case-centre scheme while considerably reducing the amount of data for delivery.  

\section{Conclusion}
In this study, we explore the application of federated split learning in LEO satellite networks, considering the constraints of limited computing resources. We propose SL-LEO, an asynchronous algorithm specifically designed to enhance communication efficiency. Furthermore, SL-LEO is capable of adapting to the model split in scenarios where satellites possess different computing capacities. Through extensive simulations, we demonstrate that SL-LEO outperforms baseline algorithms in terms of accuracy improvement for both IID and non-IID datasets. Our findings emphasize the effectiveness of SL-LEO in improving the performance of federated split learning in LEO satellite networks with limited computing resources. As a potential future direction, we are looking forward to extending our method to improve the performance of various applications such as large language models~\cite{lin2024splitlora,fang2024automated,wang2025contemporary,hu2024agentscodriver,qiu2024ifvit,hu2024agentscomerge,zhou2025survey,lin2023pushing}, reinforcement learning system~\cite{zhang2025robust,duan2025rethinking,zhang2025state}, and multimodal training~\cite{tang2024merit,fang2024ic3m}.

\ifCLASSOPTIONcaptionsoff
  \newpage
\fi

\bibliographystyle{IEEEtran}
\bibliography{reference}

\end{document}